\documentclass[english,english,12pt,onecolumn, draftcls]{IEEEtran}
\usepackage[T1]{fontenc}
\usepackage[latin9]{inputenc}
\usepackage{mathrsfs}
\usepackage{amsthm}
\usepackage{amsmath}
\usepackage{graphicx}
\PassOptionsToPackage{version=3}{mhchem}
\usepackage{mhchem}

\makeatletter
\theoremstyle{plain}
\newtheorem{thm}{\protect\theoremname}
\theoremstyle{definition}
\newtheorem{defn}[thm]{\protect\definitionname}
\theoremstyle{plain}
\newtheorem{prop}[thm]{\protect\propositionname}
\theoremstyle{remark}
\newtheorem{rem}[thm]{\protect\remarkname}

\ifCLASSINFOpdf
\else
\fi
\theoremstyle{plain}
\newtheorem{propo}{Proposition}
\renewenvironment{prop}{\begin{propo}}{\end{propo}}
\theoremstyle{remark}
\newtheorem{rema}{Remark}
\renewenvironment{rem}{\begin{rema}}{\end{rema}}

\usepackage{babel}

\providecommand{\definitionname}{Definition}
\providecommand{\propositionname}{Proposition}
\providecommand{\remarkname}{Remark}
\providecommand{\theoremname}{Theorem}

\makeatother

\usepackage{babel}
\providecommand{\definitionname}{Definition}
\providecommand{\propositionname}{Proposition}
\providecommand{\remarkname}{Remark}
\providecommand{\theoremname}{Theorem}

\begin{document}

\title{Heegard-Berger and Cascade Source Coding Problems with Common Reconstruction
Constraints}

\author{Behzad~Ahmadi,~\IEEEmembership{Student Member,~IEEE,} Ravi Tandon,~\IEEEmembership{Member,~IEEE,}
Osvaldo Simeone,~\IEEEmembership{Member,~IEEE,} H. Vincent Poor,~\IEEEmembership{Fellow,~IEEE}
\thanks{B. Ahmadi and O. Simeone are with the CWCSPR, New Jersey Institute
of Technology, Newark, NJ 07102 USA (e-mail: \{behzad.ahmadi,osvaldo.simeone\}@njit.edu). %
}
\thanks{R. Tandon was with Princeton University, Princeton, NJ 08544 USA.
He is now with Virginia Tech, Blacksburg, VA, 24061 (e-mail: tandonr@vt.edu,
rtandon@princeton.edu). %
}%
\thanks{H. V. Poor is with Princeton University, Princeton, NJ 08544 USA (e-mail:
poor@princeton.edu).%
}%
\thanks{The material in this paper was presented in part at the IEEE International
Symposium on Information Theory, Cambridge, MA, USA, July 1-6, 2012.%
}}
\maketitle
\begin{abstract}
In lossy source coding with side information at the decoder (i.e.,
the Wyner-Ziv problem), the estimate of the source obtained at the
decoder cannot be generally reproduced at the encoder, due to its
dependence on the side information. In some applications this may
be undesirable, and a Common Reconstruction (CR) requirement, whereby
one imposes that the encoder and decoder be able to agree on the decoder's
estimate, may be instead in order. The rate-distortion function under
the CR constraint has been recently derived for a point-to-point (Wyner-Ziv)
problem. In this paper, this result is extended to three multiterminal
settings with three nodes, namely the Heegard-Berger (HB) problem,
its variant with cooperating decoders and the cascade source coding
problem. The HB problem consists of an encoder broadcasting to two
decoders with respective side information. The cascade source coding
problem is characterized by a two-hop system with side information
available at the intermediate and final nodes. 

For the HB problem with the CR constraint, the rate-distortion function
is derived under the assumption that the side information sequences
are (stochastically) degraded. The rate-distortion function is also
calculated explicitly for three examples, namely Gaussian source and
side information with quadratic distortion metric, and binary source
and side information with erasure and Hamming distortion metrics.
The rate-distortion function is then characterized for the HB problem
with cooperating decoders and (physically) degraded side information.
For the cascade problem with the CR constraint, the rate-distortion
region is obtained under the assumption that side information at the
final node is physically degraded with respect to that at the intermediate
node. For the latter two cases, it is worth emphasizing that the corresponding
problem without the CR constraint is still open. Outer and inner bounds
on the rate-distortion region are also obtained for the cascade problem
under the assumption that the side information at the intermediate
node is physically degraded with respect to that at the final node.
For the three examples mentioned above, the bounds are shown to coincide.
Finally, for the HB problem, the rate-distortion function is obtained
under the more general requirement of constrained reconstruction,
whereby the decoder's estimate must be recovered at the encoder only
within some distortion.
\end{abstract}


\IEEEpeerreviewmaketitle

\section{Introduction}

\begin{figure}[h!]
\centering \includegraphics[bb=77bp 515bp 415bp 665bp,clip,scale=0.7]{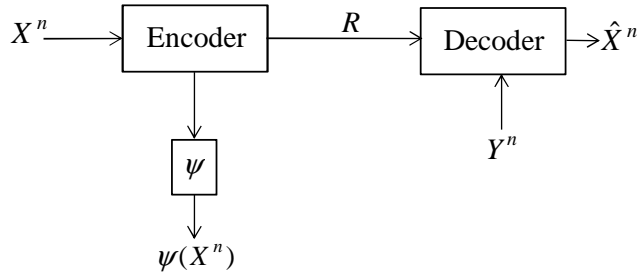}
\caption{Point-to-point source coding with common reconstruction \cite{Steinberg}.}

\label{fig:fig1}
\end{figure}

Source coding problems with side information at the decoder(s) model
a large number of scenarios of practical interest, including video
streaming \cite{Girod} and wireless sensor networks \cite{Liveris}.
From an information theoretic perspective, the baseline setting for
this class of problems is one in which a memoryless source $X^{n}=(X_{1},...,X_{n})$
is to be communicated by an encoder at a rate $R$ bits per source
symbol to a decoder that has available a correlated sequence $Y^{n}$
that is related to $X^{n}$ via a memoryless channel $p(y|x)$ (see
Fig. \ref{fig:fig1}%
\footnote{The presence of the function $\psi$ at the encoder will be explained
later.%
}). Under the requirement of asymptotically lossless reconstruction
$\hat{X}^{n}$ of the source $X^{n}$ at the decoder, the minimum
required rate was obtained by Slepian and Wolf in \cite{Slepian Wolf}.
Later, the more general optimal trade-off between rate $R$ and the
distortion $D$ between the source $X^{n}$ and reconstruction $\hat{X}^{n}$
was obtained by Wyner and Ziv in \cite{Wyner Ziv} for any given distortion
metric $d(x,\hat{x})$. It was shown to be given by the rate-distortion
function 
\begin{eqnarray}
R_{X|Y}^{WZ}(D) & =\textrm{min} & I(X;U|Y),\label{eq:RWZ}
\end{eqnarray}
where the minimum is taken over all probability mass functions (pmfs)
$p(u|x)$ and deterministic function $\mathrm{\hat{x}}(u,y)$ such
that $\mbox{E}[d(X,$ $\mathrm{\hat{x}}(U,Y))]\leq D$. 

\subsection{Heegard-Berger and Cascade Source Coding Problems}

\begin{figure}[h!]
\centering \includegraphics[bb=53bp 500bp 415bp 715bp,clip,scale=0.7]{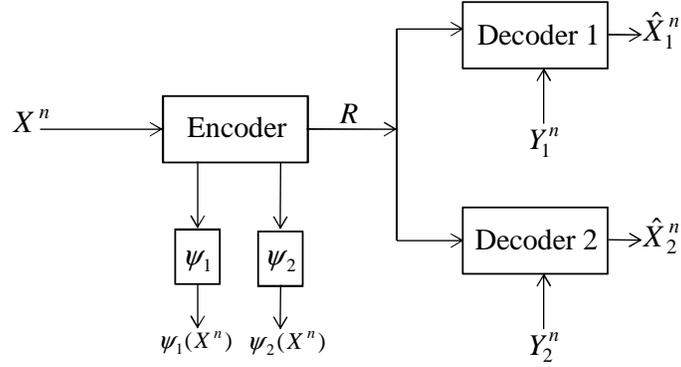}
\caption{Heegard-Berger source coding problem with common reconstruction.}

\label{fig:fig2}
\end{figure}

\begin{figure}[h!]
\centering \includegraphics[bb=54bp 495bp 420bp 752bp,clip,scale=0.7]{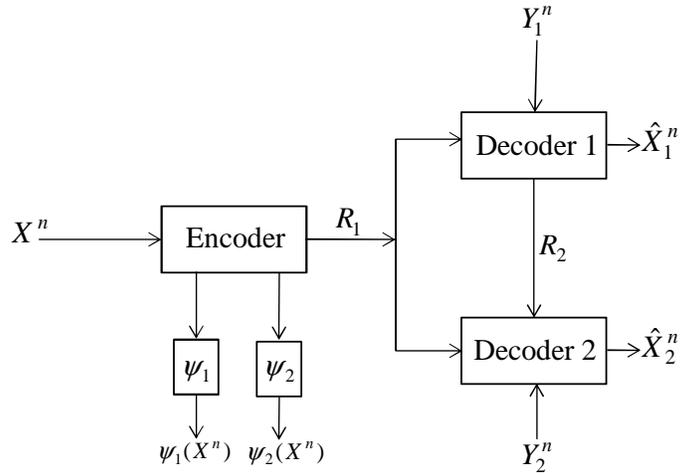}
\caption{Heegard-Berger source coding problem with common reconstruction and
decoder cooperation.}

\label{fig:fig2_cop}
\end{figure}

In applications such as the ones discussed above, the point-to-point
setting of Fig. \ref{fig:fig1} does not fully capture the main features
of the source coding problem. For instance, in video streaming, a
transmitter typically \emph{broadcasts} information to a number of
decoders. As another example, in sensor networks, data is typically
routed over \emph{multiple hops} towards the destination. A model
that accounts for the aspect of \emph{broadcasting} to multiple decoders
is the \emph{Heegard-Berger} \emph{(HB)} set-up shown in Fig. \ref{fig:fig2}.
In this model, the link of rate $R$ bits per source symbol is used
to communicate to two receivers having different side information
sequences, $Y_{1}^{n}$ and $Y_{2}^{n}$, which are related to source
$X^{n}$ via a memoryless channel $p(y_{1},y_{2}|x)$. The set of
all achievable triples ($R,D_{1},D_{2}$) for this model, where $D_{1}$
and $D_{2}$ are the distortion levels at Decoders 1 and 2 respectively,
was derived in \cite{HB} and \cite{Kaspi} under the assumption that
the side information sequences are (stochastically) degraded versions
of the source $X^{n}$. In a variation of this model shown in Fig.
\ref{fig:fig2_cop}, decoder cooperation is enabled by a limited capacity
link from one decoder (Decoder 1) to the other (Decoder 2). Inner
and outer bounds to the rate distortion region for this problem are
obtained in \cite{Vasudevan_cop} under the assumption that the side
information of Decoder 2 is (physically) degraded with respect to
that of Decoder 1.

\begin{figure}[h!]
\centering \includegraphics[bb=48bp 480bp 515bp 666bp,clip,scale=0.7]{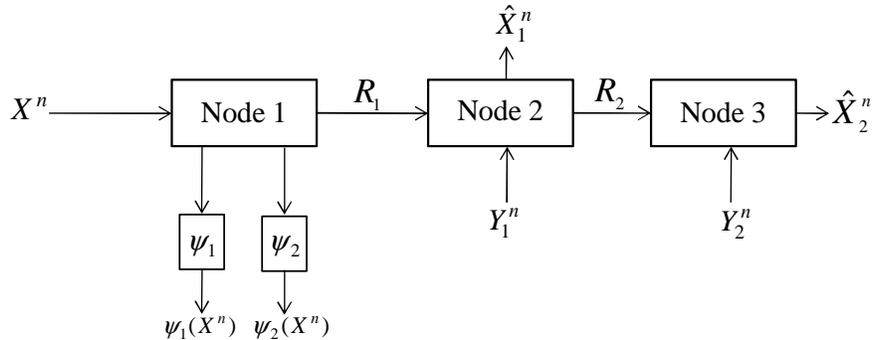}
\caption{Cascade source coding problem with common reconstruction.}

\label{fig:cascade}
\end{figure}

As for \emph{multihopping}, a basic model that captures some of the
key design issues is shown in Fig. \ref{fig:cascade}. In this \emph{cascade}
set-up, an encoder (Node 1) communicates with rate $R_{1}$ to a intermediate
node (Node 2), which has side information $Y_{1}^{n}$, and in turns
communicates with rate $R_{2}$ to a final node (Node 3) with side
information $Y_{2}^{n}$. Both Node 2 and Node 3 act as decoders,
similar to the HB problem of Fig. \ref{fig:fig2}, in the sense that
they reconstruct a local estimate of the source $X^{n}$. The rate-distortion
function for this problem has been derived for various special cases
in \cite{Yamamoto,Vasudevan,Chia} and \cite{Ravi} (see Table I in
\cite{Ravi} for an overview). Reference \cite{Chia} derives the
set of all achievable quadruples $(R_{1},R_{2},D_{1},D_{2})$, i.e.,
the rate-distortion region, for the case in which $Y_{1}^{n}$ is
also available at the encoder and $Y_{2}^{n}$ is a physically degraded
version of $X^{n}$ with respect to $Y_{1}^{n}$. Instead, \cite{Vasudevan}
derives the rate-distortion region under the assumptions that the
source and the side information sequences are jointly Gaussian, that
the distortion metric is quadratic, and that the sequence $Y_{1}^{n}$
is a physically degraded version of $X^{n}$ with respect to $Y_{2}^{n}$.
The corresponding result for binary source and side information and
Hamming distortion metric was derived in \cite{Ravi}. 

\subsection{Common Reconstruction Constraint}

A key aspect of the optimal strategies identified in \cite{Wyner Ziv,HB,Kaspi,Vasudevan}
and \cite{Chia} is that the side information sequences are, in general,
used in two different ways: (\emph{i}) as a means to reduce the rate
required for communication between encoder and decoders via binning;
and (\emph{ii}) as an additional observation that the decoder can
leverage, along with the bits received from the encoder, in order
to improve its local estimate. For instance, for the point-to-point
system of Fig. \ref{fig:fig1}, the Wyner-Ziv result (\ref{eq:RWZ})
reflects point (\emph{i}) of the discussion above in the conditioning
on side information $Y$, which reduces the rate, and point (\emph{ii})
in the fact that the reconstruction $\hat{X}$ is a function $\mathrm{\hat{x}}(U,Y)$
of the signal $U$ received from the encoder and the side information
$Y$. 

Leveraging the side information as per point (\emph{ii}), while advantageous
in terms of rate-distortion trade-off, may have unacceptable consequences
for some applications. In fact, this use of side information entails
that the reconstruction $\hat{X}$ of the decoder cannot be reproduced
at the encoder. In other words, encoder and decoder cannot agree on
the specific reconstruction $\hat{X}$ obtained at the receiver side,
but only on the average distortion level $D$. In applications such
as transmission of sensitive medical, military or financial data,
this may not be desirable. Instead, one may want to add the constraint
that the reconstruction at the decoder be reproducible by the encoder
\cite{Steinberg}. This idea, referred to as the Common Reconstruction
(CR) constraint, was first proposed in \cite{Steinberg}, where it
is shown for the point-to-point setting of Fig. \ref{fig:fig1}%
\footnote{The function $\psi$ at the encoder calculates the estimate of the
encoder regarding the decoder's reconstruction.%
} that the rate-distortion function under the CR constraint is given
by
\begin{eqnarray}
R_{X|Y}^{CR}(D) & =\textrm{min} & I(X;\hat{X}|Y),\label{eq:RCR}
\end{eqnarray}
where the minimum is taken over all pmfs $p(\hat{x}|x)$ such that
$\mbox{E}[d(X,\hat{X})]\leq D$. Comparing (\ref{eq:RCR}) with the
Wyner-Ziv rate-distortion (\ref{eq:RWZ}), it can be seen that the
additional CR constraint prevents the decoder from using the side
information as a means to improve its estimate $\hat{X}$ (see point
(\emph{ii}) above).

The original work of \cite{Steinberg} has been recently extended
in \cite{Lapidoth}, where a relaxed CR constraint is imposed in which
only a distortion constraint is imposed between the decoder's reconstruction
and its reproduction at the encoder. We refer to this setting as imposing
a \emph{Constrained Reconstruction} (ConR) requirement.

\subsection{Main Contributions}

In this paper, we study the HB source coding problem (Fig. \ref{fig:fig2})
and the cascade source coding problem (Fig. \ref{fig:cascade}) under
the CR requirement. The considered models are thus relevant for the
transmission of sensitive information, which is constrained by CR,
via broadcast or multi-hop links \textendash{} a common occurrence
in, e.g., medical, military or financial applications (e.g., for intranets
of hospitals or financial institutions). Specifically, our main contributions
are:
\begin{itemize}
\item For the HB problem with the CR constraint (Fig. \ref{fig:fig2}),
we derive the rate-distortion function under the assumption that the
side information sequences are (stochastically) degraded. We also
calculate this function explicitly for three examples, namely Gaussian
source and side information with quadratic distortion metric, and
binary source and erasure side information with erasure and Hamming
distortion metrics (Sec. \ref{sec:HB});
\item For the HB problem with the CR constraint and decoder cooperation
(Fig. \ref{fig:fig2_cop}), we derive the rate-distortion region under
the assumption that the side information sequences are (physically)
degraded in either direction (Sec. \ref{sub:xy1y2_cop} and Sec. \ref{sub:xy2y1_cop}).
We emphasize that the corresponding problem without the CR constraint
is still open as per the discussion above; 
\item For the cascade problem with the CR constraint (Fig. \ref{fig:cascade}),
we obtain the rate-distortion region under the assumption that side
information $Y_{2}$ is physically degraded with respect to $Y_{1}$
(Sec. \ref{sub:xy1y2}). We emphasize that the corresponding problem
without the CR constraint is still open as per the discussion above;
\item For the cascade problem with CR constraint (Fig. \ref{fig:cascade}),
we obtain outer and inner bounds on the rate-distortion region under
the assumption that the side information $Y_{1}$ is physically degraded
with respect to $Y_{2}$. Moreover, for the three examples mentioned
above in the context of the HB problem, we show that the bounds coincide
and we evaluate the corresponding rate-distortion region explicitly
(Sec. \ref{sub:xy2y1});
\item For the HB problem, we finally derive the rate-distortion function
under the more general requirement of ConR (Sec. \ref{sec:ConR}).
\end{itemize}
\textit{Notation}: For $a$ and $b$ integer with $a\leq b$ we define
$[a,b]$ as the interval $[a,a+1,...,b]$ and we use $x_{a}^{b}$
to denote the sequence $(x_{a},\ldots,x_{b})$. We will also write
$x^{b}$ for $x_{1}^{b}$ for simplicity. Upper case, lower case and
calligraphic letters denote random variables, specific values of random
variables and their alphabets, respectively. Given discrete random
variables, or more generally vectors, $X$ and $Y$, we will use the
notation $p_{X}(x)$ or $p(x)$ for $\ce{Pr}[X=x]$, and $p_{X|Y}(x|y)$
or $p(x|y)$ for $\ce{Pr}[X=x|Y=y]$, where the latter notations are
used when the meaning is clear from the context. Given a set $\mathcal{X}$,
we denoted by $\mathcal{X}^{n}$ the $n$-fold Cartesian product of
$\mathcal{X}$. For random variables $X$ and $Y$, we denote by $\sigma_{X|Y}^{2}$
the (average) conditional variance of $X$ given $Y$, i.e., $\textrm{E}\left[\textrm{E}[(X-\textrm{E}[X|Y])^{2}|Y]\right].$
We adopt the notation convention in \cite{Elgammal}, in which $\delta(\epsilon)$
represents any function such that $\delta(\epsilon)\rightarrow0$
as $\epsilon\rightarrow0$. We define the binary entropy function
$H(p)=-p\textrm{lo\ensuremath{g_{2}}}p-(1-p)\textrm{lo\ensuremath{g_{2}}}(1-p)$.
Finally, we define $\alpha*\beta=\alpha(1-\beta)+\beta(1-\alpha)$.

\section{Heegard-Berger Problem with Common Reconstruction\label{sec:HB}}

In this section, we first detail the system model for the HB source
coding problem in Fig. \ref{fig:fig2} with CR in Sec. \ref{sub:System-model_HB}.
Next, the characterization of the corresponding rate-distortion performance
is derived under the assumption that one of the two side information
sequences is a stochastically degraded version of the other in the
sense of \cite{HB} (see (\ref{eq:degraded})). Finally, three specific
examples are worked out, namely Gaussian sources under quadratic distortion
(Sec. \ref{sub:Gaussian-HB}), and binary sources with side information
sequences subject to erasures under Hamming or erasure distortion
(Sec. \ref{sub:binary-HB}). 



\subsection{System Model\label{sub:System-model_HB}}

In this section the system model for the HB problem with CR is detailed.
The system is defined by the pmf $p_{XY_{1}Y_{2}}(x,y_{1},y_{2})$
and discrete alphabets $\mathcal{X},\mathcal{Y}_{1},\mathcal{Y}_{2},\hat{\mathcal{X}}_{1},$
and $\hat{\mathcal{X}}_{2}$ as follows. The source sequence $X^{n}$
and side information sequences $Y_{1}^{n}$ and $Y_{2}^{n}$, with
$X^{n}\in\mathcal{X}^{n}$, $Y_{1}^{n}\in\mathcal{Y}_{1}^{n}$, and
$Y_{2}^{n}\in\mathcal{Y}_{2}^{n}$ are such that the tuples $(X_{i},Y_{1i},Y_{2i})$
for $i\in[1,n]$ are independent and identically distributed (i.i.d.)
with joint pmf $p_{XY_{1}Y_{2}}(x,y_{1},y_{2})$. The encoder measures
a sequence $X^{n}$ and encodes it into a message $J$ of $nR$ bits,
which is delivered to the decoders. Decoders 1 and 2 wish to reconstruct
the source sequence $X^{n}$ within given distortion requirements,
to be discussed below, as $\hat{X}_{1}^{n}\in\mathcal{\hat{X}}_{1}^{n}$
and $\hat{X}_{2}^{n}\in\mathcal{\hat{X}}_{2}^{n}$, respectively.
The estimated sequence $\hat{X}_{j}^{n}$ is obtained as a function
of the message $J$ and the side information sequence $Y_{j}^{n}$
for $j=1,2$. The estimates are constrained to satisfy distortion
constraints defined by per-symbol distortion metrics $d_{j}(x,\hat{x}_{j}):\mathcal{X}\times\mathcal{\hat{X}}_{j}\rightarrow[0,D_{max}]$
with $0<D_{max}<\infty$. Based on the given distortion metrics, the
overall distortion for the estimated sequences $\hat{x}_{1}^{n}$
and $\hat{x}_{2}^{n}$ is defined as
\begin{align}
d_{j}^{n}(x^{n},\hat{x}_{j}^{n}) & =\frac{1}{n}\sum_{i=1}^{n}d_{j}(x_{i},\hat{x}_{ji})\mbox{ for \ensuremath{j=1,2.}}\label{eq:overall_dist}
\end{align}
The reconstructions $\hat{X}_{2}^{n}$ and $\hat{X}_{2}^{n}$ are
also required to satisfy the CR constraints, as formalized below.
\begin{defn}
\label{HBcode_def}An $(n,R,D_{1},D_{2},\epsilon)$ code for the HB
problem with CR consists of an encoding function
\begin{align}
g\textrm{: }\mathcal{X}^{n}\rightarrow[1,2^{nR}],\label{eq:enc}
\end{align}
which maps the source sequence $X^{n}$ into a message $\mbox{\ensuremath{J};}$
a decoding function for Decoder 1,
\begin{align}
h_{1}\textrm{: }[1,2^{nR}]\times\mathcal{Y}_{1}^{n}\rightarrow\mathcal{\hat{X}}_{1}^{n},\label{eq:dec1}
\end{align}
which maps the message $J$ and the side information $Y_{1}^{n}$
into the estimated sequence $\hat{X}_{1}^{n}$; a decoding function
for Decoder 2
\begin{align}
h_{2}\textrm{: }[1,2^{nR}]\times\mathcal{Y}_{2}^{n} & \rightarrow\mathcal{\hat{X}}_{2}^{n}\label{eq:dec2}
\end{align}
which maps message $J$ and the side information $Y_{2}^{n}$ into
the estimated sequence $\hat{X}_{2}^{n}$; and two reconstruction
functions\begin{subequations}\label{eqn: en_recons}
\begin{align}
 & \psi_{1}\textrm{: }\mathcal{X}^{n}\rightarrow\mathcal{\hat{X}}_{1}^{n}\\
\textrm{and } & \psi_{2}\textrm{: }\mathcal{X}^{n}\rightarrow\mathcal{\hat{X}}_{2}^{n},
\end{align}
\end{subequations}which map the source sequence into the estimated
sequences at the encoder, namely $\psi_{1}(X^{n})$ and $\psi_{2}(X^{n})$,
respectively; such that the distortion constraints are satisfied,
i.e., 
\begin{eqnarray}
\frac{1}{n}\sum_{i=1}^{n}\textrm{E}\left[d_{j}(X_{i},\hat{X}_{ji})\right] & \leq & D_{j}\mbox{ for \ensuremath{j=1,2}},\label{eq:dist_const}
\end{eqnarray}
and the CR requirements hold, namely, 
\begin{eqnarray}
\textrm{Pr}\left[\psi_{j}(X^{n})\neq\hat{X}_{j}^{n}\right] & \leq & \epsilon,\textrm{ }j=1,2.\label{eq:CR_req}
\end{eqnarray}
Given distortion pairs $(D_{1},D_{2})$, a rate pair $R$ is said
to be achievable if, for any $\epsilon>0$ and sufficiently large
$n$, there exists an $(n,R,D_{1}+\epsilon,D_{2}+\epsilon,\epsilon)$
code. The rate-distortion function $R(D_{1},D_{2})$ is defined as
$R(D_{1},D_{2})=$inf$\{R:$ the triple $(R,D_{1},D_{2})$ is achievable\}.
\end{defn}

\subsection{Rate-Distortion Function}

In this section, a single-letter characterization of the rate-distortion
function for the HB problem with CR is derived, under the assumption
that the joint pmf $p(x,y_{1},y_{2})$ is such that there exists a
conditional pmf $\tilde{p}(y_{1}|y_{2})$ for which
\begin{align}
p(x,y_{1}) & =\overset{}{\underset{y_{2}\in\mathcal{Y}_{2}}{\sum}}p(x,y_{2})\tilde{p}(y_{1}|y_{2}).\label{eq:degraded}
\end{align}
In other words, the side information $Y_{1}$ is a stochastically
degraded version of $Y_{2}.$ 
\begin{prop}
\label{prop:HB_converse}If the side information $Y_{1}$ is stochastically
degraded with respect to $Y_{2}$, the rate-distortion function for
the HB problem with CR is given by
\begin{eqnarray}
R_{HB}^{CR}(D_{1},D_{2}) & = & \min I(X;\hat{X}_{1}|Y_{1})+I(X;\hat{X}_{2}|Y_{2}\hat{X}_{1})\label{eq:RHB}
\end{eqnarray}
where the mutual information terms are evaluated with respect to the
joint pmf
\begin{align}
p(x,y_{1},y_{2},\hat{x}_{1},\hat{x}_{2})=p(x,y_{1},y_{2})p(\hat{x}_{1},\hat{x}_{2}|x) & ,\label{eq:joint}
\end{align}
and minimization is performed with respect to the conditional pmf
$p(\hat{x}_{1},\hat{x}_{2}|x)$ under the constraints\textup{
\begin{align}
\mathrm{E}[d_{j}(X,\hat{X}_{j})]\leq D_{j},\mbox{ for }j=1,2.\label{eq:const}
\end{align}
}
\end{prop}
The proof of the converse can be found in Appendix A. Achievability
follows as a special case of Theorem 3 of \cite{HB} and can be easily
shown using standard arguments. In particular, the encoder randomly
generates a standard lossy source code $\hat{X}_{1}^{n}$ for the
source $X^{n}$ with rate $I(X;\hat{X}_{1})$ bits per source symbol.
Random binning is used to reduce the rate to $I(X;\hat{X}_{1}|Y_{1}).$
By the Wyner-Ziv theorem \cite[p. 280]{Elgammal}, this guarantees
that both Decoder 1 and Decoder 2 are able to recover $\hat{X}_{1}^{n}$
(since $Y_{1}$ is a degraded version of $Y_{2}$). The encoder then
maps the source $X^{n}$ into the reconstruction sequence $\hat{X}_{2}^{n}$
using a codebook that is generated conditional on $\hat{X}_{1}^{n}$
with rate $I(X;\hat{X}_{2}|\hat{X}_{1})$ bits per source symbol.
Random binning is again used to reduce the rate to $I(X;\hat{X}_{2}|Y_{2}\hat{X}_{1})$.
From the Wyner-Ziv theorem, and the fact that Decoder 2 knows the
sequence $\hat{X}_{1}^{n}$, it follows that Decoder 2 can recover
the reconstruction $\hat{X}_{2}^{n}$ as well. Note that, since the
reconstruction sequences $\hat{X}_{1}^{n}$ and $\hat{X}_{2}^{n}$
are generated by the encoder, functions $\psi_{1}$ and $\psi_{2}$
that guarantees the CR constraints (\ref{eq:CR_req}) exist by construction.

\begin{rem}Under the physical degradedness assumption that the Markov
chain condition $X\textrm{---}Y_{2}\textrm{---}Y_{1}$ holds, equation
(\ref{eq:RHB}) can be rewritten as
\begin{eqnarray}
R=\min I(X;\hat{X}_{1}\hat{X}_{2}|Y_{2})+I(\hat{X}_{1};Y_{2}|Y_{1}),\label{eq:RHB1}
\end{eqnarray}
with the minimization defined as in (\ref{eq:RHB}). This expression
quantifies by $I(\hat{X}_{1};Y_{2}|Y_{1})$ the additional rate that
is required with respect to the ideal case in which both decoders
have the better side information $Y_{2}$.

\end{rem}
\begin{rem}
\label{HB}If we remove the CR constraint, then the rate-distortion
function under the assumption of Proposition \ref{prop:HB_converse}
is given by \cite{HB}
\begin{eqnarray}
R_{HB}(D_{1},D_{2}) & = & \min I(X;U_{1}|Y_{1})+I(X;U_{2}|Y_{2}U_{1}),\label{eq:RHB-1}
\end{eqnarray}
where the mutual information terms are evaluated with respect to the
joint pmf
\begin{align}
p(x,y_{1},y_{2},u_{1},u_{2},\hat{x}_{1},\hat{x}_{2})=p(x,y_{1},y_{2})p(u_{1},u_{2}|x)\delta(\hat{x}_{1}-\mathrm{\hat{x}_{1}}(u_{1},y_{1}))\delta(\hat{x}_{2}-\mathrm{\hat{x}_{2}}(u_{2},y_{2})) & ,\label{eq:joint-1}
\end{align}
and minimization is performed with respect to the conditional pmf
$p(u_{1},u_{2}|x)$ and the deterministic functions $\mathrm{\mathrm{\hat{x}_{j}}\mbox{\ensuremath{(u_{j},y_{j})}}},$
for $j=1,2$, such that distortion constraints (\ref{eq:const}) are
satisfied. Comparison of (\ref{eq:RHB}) with (\ref{eq:RHB-1}) reveals
that, similar to the discussion around (\ref{eq:RWZ}) and (\ref{eq:RCR}),
the CR constraint permits the use of side information only to reduce
the rate via binning, but not to improve the decoder's estimates via
the use of the auxiliary codebooks represented by variables $U_{1}$
and $U_{2}$, and functions $\mathrm{\mathrm{\hat{x}_{j}}\mbox{\ensuremath{(u_{j},y_{j})}}},$
for $j=1,2$, in (\ref{eq:joint-1}).

\begin{rem}\label{causal-0}Consider the case in which the side information
sequences are available in a causal fashion in the sense of \cite{Weissman},
that is, the decoding functions (\ref{eq:dec1})-(\ref{eq:dec2})
are modified as $h_{ji}\textrm{: }[1,2^{nR}]\times\mathcal{Y}_{j}^{i}\rightarrow\mathcal{\hat{X}}_{ji}$,
for $i\in[1,n]$ and $j=1,2$, respectively. Following similar steps
as in the proof of Proposition 2 and in \cite{Weissman}, it can be
concluded that, under the CR constraint, the rate-distortion function
in this case is the same as if the two side information sequences
were not available at the decoders, and is thus given by (\ref{eq:RHB})
upon removing the conditioning on the side information. Note that
this is true irrespective of the joint pmf $p(x,y_{1},y_{2})$ and
hence it holds also for non-degraded side information. This result
can be explained by noting that, as explained in \cite{Weissman},
causal side information prevents the possibility of reducing the rate
via binning. Since the CR constraint also prevents the side information
from being used to improve the decoders' estimates, it follows that
the side information is useless in terms of rate-distortion performance,
if used causally under the CR constraint. 

On a similar note, if only side information $Y_{1}$ is causally available,
while $Y_{2}$ can still be used in the conventional non-causal fashion,
then it can be proved that $Y_{1}$ can be neglected without loss
of optimality. Therefore, the rate-distortion function follows from
(\ref{eq:RHB}) by removing the conditioning on $Y_{1}$.
\end{rem}
\end{rem}

\begin{rem}In \cite{Timo}, a related model is studied in which the
source is given as $X=(Y_{1},Y_{2})$ and each decoder is interested
in reconstructing a lossy version of the side information available
at the other decoder. The CR constraint is imposed in a different
way by requiring that each decoder be able to reproduce the estimate
reconstructed at the other decoder.\end{rem}

\subsection{Gaussian Sources and Quadratic Distortion\label{sub:Gaussian-HB} }

In this section, we highlight the result of Proposition \ref{prop:HB_converse}
by considering a zero-mean Gaussian source $X\sim\mathcal{N}(0,\sigma_{x}^{2})$,
with side information variables\begin{subequations}\label{eqn:side_Gauss}
\begin{eqnarray}
Y_{1} & = & X+Z_{1}\\
\textrm{and }Y_{2} & = & X+Z_{2},
\end{eqnarray}
\end{subequations}where $Z_{1}\sim\mathcal{N}(0,N_{1}+N_{2})$ and
$Z_{2}\sim\mathcal{N}(0,N_{2})$ are independent of each other and
of $Y_{2}$ and $X$. Note that the joint distribution of $(X,Y_{1},Y_{2})$
satisfies the stochastic degradedness condition. We focus on the quadratic
distortion $d_{j}(x,\hat{x}_{j})=(x-\hat{x}_{j})^{2}$ for $j=1,2$.
By leveraging standard
\begin{figure}[h!]
\centering \includegraphics[bb=78bp 366bp 482bp 688bp,clip,scale=0.7]{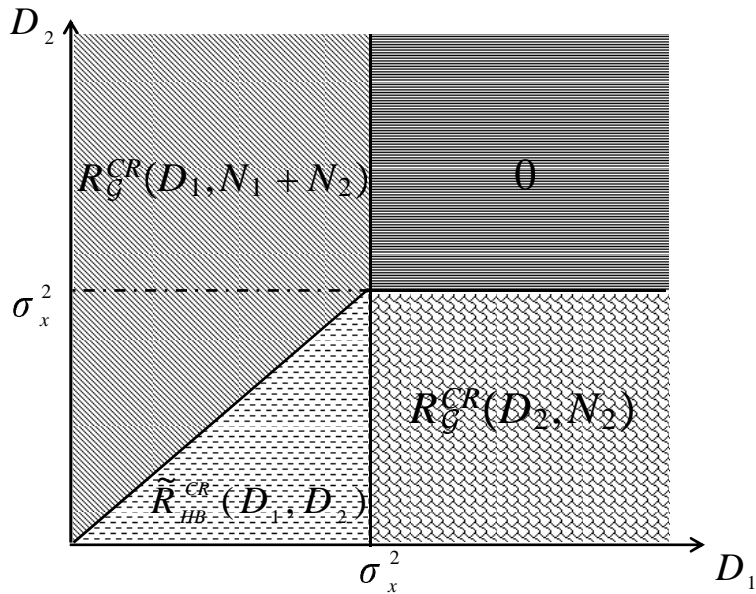}
\caption{Illustration of the distortion regions in the rate-distortion function
(\ref{eq:region}) for Gaussian sources and quadratic distortion.}

\label{fig:region}
\end{figure}
arguments that allow us to apply Proposition \ref{prop:HB_converse}
to Gaussian sources under mean-square-error constraint (see \cite[pp. 50-51]{Elgammal}
and \cite{Wyner}), we obtain a characterization of the rate-distortion
function for the given distortion and metrics.

We first recall that for the point-to-point set-up in Fig. \ref{fig:fig1}
with $X\sim\mathcal{N}(0,\sigma_{x}^{2})$ and side information $Y=X+Z,$
with $Z\sim\mathcal{N}(0,N)$ independent of $X,$ the rate-distortion
function with CR under quadratic distortion is given by \cite{Steinberg}
\begin{align}
R_{X|Y}^{CR}(D) & = & \left\{ \begin{array}{lc}
R_{\mathcal{G}}^{CR}(D,N)\overset{\triangle}{=}\frac{1}{2}\log_{2}\left(\frac{\sigma_{x}^{2}}{\sigma_{x}^{2}+N}\cdot\frac{D+N}{D}\right) & \mbox{ for }D\leq\sigma_{x}^{2}\\
0 & \mbox{ for }D>\sigma_{x}^{2},
\end{array}\right.\label{eq:RCR_Gauss}
\end{align}
where we have made explicit dependence on $N$ of function $R_{\mathcal{G}}^{\mathcal{\mbox{\ensuremath{CR}}}}(D,N)$
for convenience. The rate-distortion function (\ref{eq:RCR_Gauss})
for $D\leq\sigma_{x}^{2}$ is obtained from (\ref{eq:RCR}) by choosing
the distribution $p(\hat{x}|x)$ such that $X=\hat{X}+Q$ where $Q\sim\mathcal{N}(0,D)$
is independent of $\hat{X}$.
\begin{prop}
\label{prop:HB_Gauss}The rate-distortion function for the HB problem
with CR for Gaussian sources (\ref{eqn:side_Gauss}) and quadratic
distortion is given by\textup{
\begin{eqnarray}
R_{HB}^{CR}(D_{1},D_{2}) & = & \left\{ \begin{array}{ll}
0 & \mbox{if }D_{1}\geq\sigma_{x}^{2}\mbox{ and }D_{2}\geq\sigma_{x}^{2},\\
R_{\mathcal{G}}^{CR}(D_{1},N_{1}+N_{2}) & \mbox{if }D_{1}\leq\sigma_{x}^{2}\mbox{ and }D_{2}\geq\min(D_{1},\sigma_{x}^{2})\\
R_{\mathcal{G}}^{CR}(D_{2},N_{2}) & \mbox{if \ensuremath{D_{1}}\ensuremath{\geq\sigma_{x}^{2}} }\mbox{and }D_{2}\leq\sigma_{x}^{2}\\
\tilde{R}_{HB}^{CR}(D_{1},D_{2}) & \mbox{if }D_{2}\leq D_{1}\leq\sigma_{x}^{2}
\end{array}\right.\label{eq:region}
\end{eqnarray}
}where $R_{\mathcal{G}}^{CR}(D,N)$ is defined in (\ref{eq:RCR_Gauss})
and\textup{
\begin{align}
\tilde{R}_{HB}^{CR}(D_{1},D_{2}) & \overset{\bigtriangleup}{=}\frac{1}{2}\log_{2}\left(\frac{\sigma_{x}^{2}}{(\sigma_{x}^{2}+N_{1}+N_{2})}\cdot\frac{(D_{1}+N_{1}+N_{2})(D_{2}+N_{2})}{(D_{1}+N_{2})D_{2}}\right).
\end{align}
}\end{prop}
\begin{rem}
The rate-distortion function for the HB problem for Gaussian sources
(\ref{eqn:side_Gauss}) without the CR constraint can be found in
\cite{HB}. Comparison with (\ref{eq:region}) confirms the performance
loss discussed in Remark \ref{HB}. 
\end{rem}
Definition of the rate distortion function (\ref{eq:region}) requires
different consideration for the four subregions of the $(D_{1},D_{2})$
plane sketched in Fig. \ref{fig:region}. In fact, for $D_{1}\geq\sigma_{x}^{2}\mbox{ and }D_{2}\geq\sigma_{x}^{2}$,
the required rate is zero, since the distortion constraints are trivially
met by setting $\hat{X_{1}}=\hat{X_{2}}=0$ in the achievable rate
(\ref{eq:RHB}). For the case $D_{1}\geq\sigma_{x}^{2}\mbox{ and }D_{2}\leq\sigma_{x}^{2}$,
it is sufficient to cater only to Decoder 2 by setting $\hat{X_{1}}=0$
and $X=\hat{X}_{2}+Q_{2}$, with $Q_{2}\sim\mathcal{N}(0,D_{2})$
independent of $\hat{X_{2}}$, in the achievable rate (\ref{eq:RHB}).
That this rate cannot be improved upon follows from the trivial converse
\begin{align}
R_{HB}^{CR}(D_{1},D_{2})\geq\textrm{max}\{R_{\mathcal{G}}^{CR}(D_{1},N_{1}+N_{2}),R_{\mathcal{G}}^{CR}(D_{2},N_{2})\},
\end{align}
which follows by cut-set arguments. The same converse suffices also
for the regime $D_{1}\leq\sigma_{x}^{2}\mbox{ and }D_{2}\geq\min(D_{1},\sigma_{x}^{2})$.
For this case, achievability follows by setting $X=\hat{X}_{1}+Q_{1}$
and $\hat{X_{1}}=\hat{X_{2}}$ in (\ref{eq:RHB}), where $Q_{1}\sim\mathcal{N}(0,D_{1})$
is independent of $\hat{X_{1}}$. In the remaining case, namely $D_{2}\leq D_{1}\leq\sigma_{x}^{2}$,
the rate-distortion function does not follow from the point-to-point
result (\ref{eq:RCR_Gauss}) as for the regimes discussed thus far.
The analysis of this case requires use of entropy-power inequality
(EPI) and can be found in Appendix B

Fig. \ref{fig:plot1} depicts the rate $R_{HB}^{CR}(D_{1},D_{2})$
in (\ref{eq:region}) versus $D_{1}$ for different values of $D_{2}$
with $\sigma_{x}^{2}=4$, $N_{1}=2$, and $N_{2}=3.$ As discussed
above, for $D_{2}=5$, which is larger than $\sigma_{x}^{2}$, $R_{HB}^{CR}(D_{1},D_{2})$
becomes zero for values of $D_{1}$ larger than $\sigma_{x}^{2}=4$,
while this is not the case for values $D_{2}<\sigma_{x}^{2}=4$.
\begin{figure}[h!]
\centering \includegraphics[bb=40bp 159bp 550bp 553bp,clip,scale=0.7]{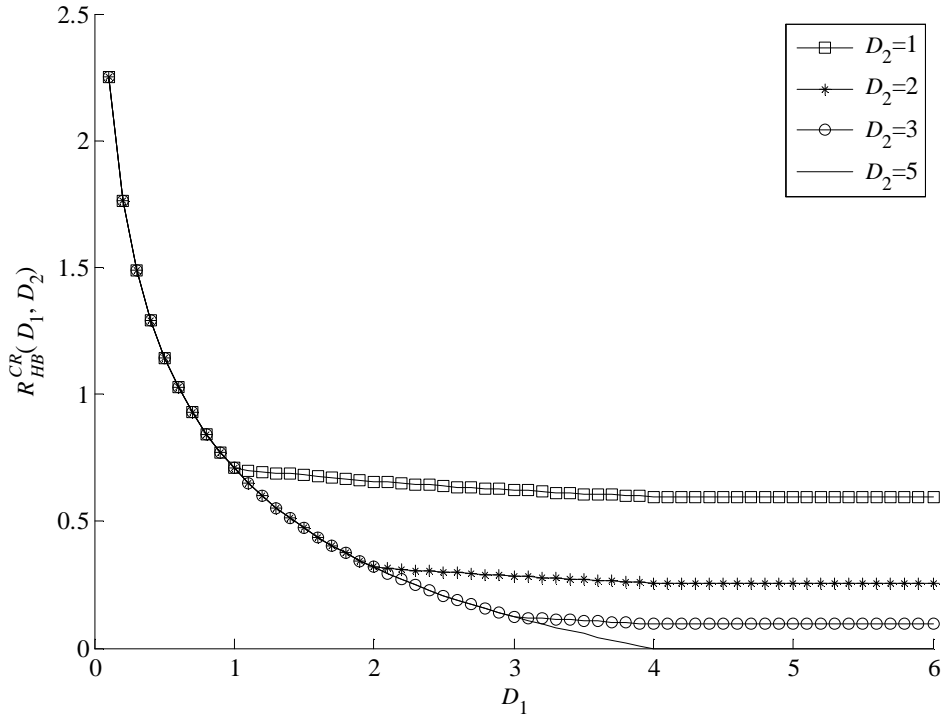}
\caption{The rate-distortion function $R_{HB}^{CR}(D_{1},D_{2})$ in (\ref{eq:region})
versus distortion $D_{1}$ for different values of distortion $D_{2}$
and for $\sigma_{x}^{2}=4$, $N_{1}=2$, and $N_{2}=3.$ }

\label{fig:plot1}
\end{figure}

\subsection{Binary Source with Erased Side Information and Hamming or Erasure
Distortion\label{sub:binary-HB}}

\begin{figure}[h!]
\centering \includegraphics[bb=152bp 433bp 455bp 600bp,clip,scale=0.6]{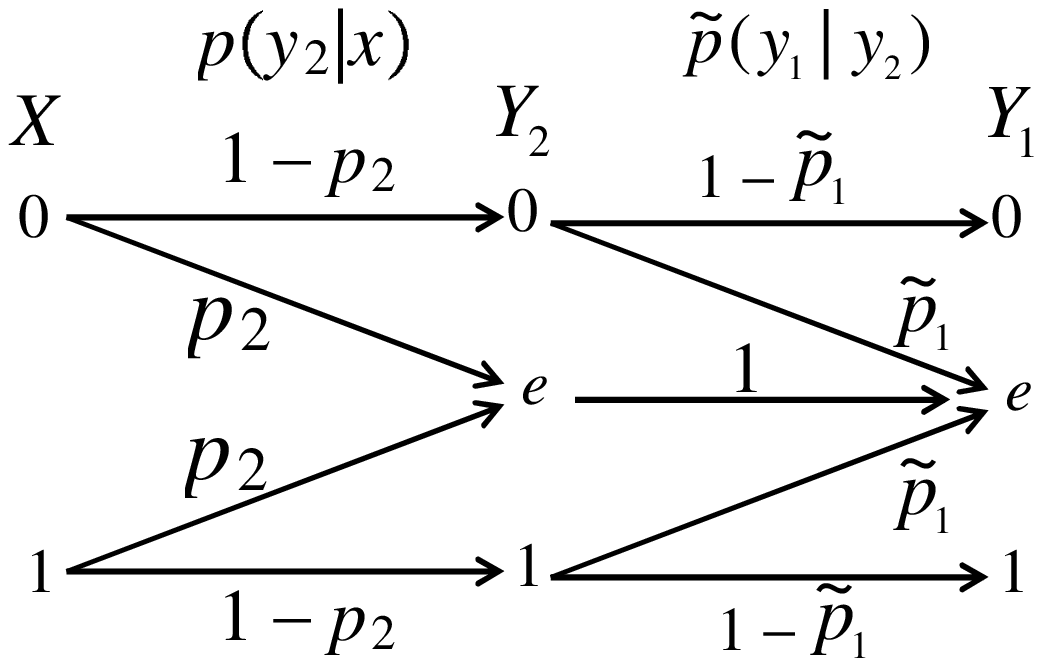}
\caption{Illustration of the pmfs in the factorization (\ref{eq:degraded})
of the joint distribution $p(x,y_{1},y_{2})$ for a binary source
$X$ and erased side information sequences $(Y_{1},Y_{2}).$ }

\label{fig:test}
\end{figure}

In this section, we consider a binary source $X\sim\textrm{Ber}(\frac{1}{2})$
with erased side information sequences $Y_{1}$ and $Y_{2}$. The
source $Y_{2}$ is an erased version of the source $X$ with erasure
probability $p_{2}$ and $Y_{1}$ is an erased version of $X$ with
erasure probability $p_{1}>p_{2}.$ This means that $Y_{j}=e$, where
$e$ represents an erasure, with probability $p_{j}$ and $Y_{j}=X$
with probability $1-p_{j}$. Note that, with these assumptions, the
side information $Y_{1}$ is stochastically degraded with respect
to $Y_{2}.$ In fact, we have the factorization (\ref{eq:degraded}),
where additional distributions $p(y_{2}|x)$ and $\tilde{p}(y_{1}|y_{2})$
are illustrated in Fig. \ref{fig:test}. As seen in Fig. \ref{fig:test},
the pmf $\tilde{p}(y_{1}|y_{2})$ is characterized by the probability
$\tilde{p}_{1}$ that satisfies the equality $p_{1}=p_{2}+\tilde{p}_{1}(1-p_{2})$.
We focus on Hamming and erasure distortions. For the Hamming distortion,
the reconstruction alphabets are binary, $\hat{\mathcal{X}}_{1}=\hat{\mathcal{X}}_{2}=\{0,1\}$,
and we have $d_{j}(x,\hat{x}_{j})=0$ if $x=\hat{x}_{j}$ and $d_{j}(x,\hat{x}_{j})=1$
otherwise for $j=1,2$. Instead, for the erasure distortion the reconstruction
alphabets are $\hat{\mathcal{X}}_{1}=\hat{\mathcal{X}}_{2}=\{0,1,e\}$,
and we have for $j=1,2$:
\begin{eqnarray}
d_{j}(x,\hat{x}_{j}) & = & \left\{ \begin{array}{lc}
0 & \mbox{ for }\hat{x}_{j}=x\\
1 & \mbox{ for }\hat{x}_{j}=e\\
\infty & \textrm{otherwise}
\end{array}\right.\label{eq:erasure_dist}
\end{eqnarray}

In Appendix C, we prove that for the point-to-point set-up in Fig.
\ref{fig:fig1} with $X\sim\textrm{Ber}(\frac{1}{2})$ and erased
side information $Y,$ with erasure probability $p$, the rate-distortion
function with CR under Hamming distortion is given by
\begin{align}
R_{X|Y}^{CR}(D) & = & \left\{ \begin{array}{lc}
R_{\mathcal{B}}^{CR}(D,p)\overset{\triangle}{=}p(1-H(D)) & \mbox{ for }D\leq1/2\\
0 & \mbox{ for }D>1/2,
\end{array}\right.\label{eq:RCR_Bin}
\end{align}
where we have made explicit the dependence on $p$ of function $R_{\mathcal{B}}^{CR}(D,p)$
for convenience. The rate-distortion function (\ref{eq:RCR_Bin})
for $D\leq1/2$ is obtained from (\ref{eq:RCR}) by choosing the distribution
$p(\hat{x}|x)$ such that $X=\hat{X}\oplus Q$ where $Q\sim\textrm{Ber}(D)$
is independent of $\hat{X}$. Following the same steps as in Appendix
C, it can be also proved that for the point-to-point set-up in Fig.
\ref{fig:fig1} with $X\sim\textrm{Ber}(\frac{1}{2})$ and erased
side information $Y,$ with erasure probability $p$, the rate-distortion
function with CR under erasure distortion is given by
\begin{eqnarray}
R_{X|Y}^{CR}(D) & = & R_{\mathcal{BE}}^{CR}(D,p)\overset{\triangle}{=}p(1-D).\label{eq:RCR_Bin-1}
\end{eqnarray}
The rate-distortion function (\ref{eq:RCR_Bin-1}) is obtained from
(\ref{eq:RCR}) by choosing the distribution $p(\hat{x}|x)$ such
that $\hat{X}=X$ with probability $1-D$ and $\hat{X}=e$ with probability
$D$.
\begin{rem}
The rate-distortion function with erased side information and Hamming
distortion without the CR constraint is derived in \cite{Perron}
(see also \cite{Verdu}). Comparison with (\ref{eq:RCR_Bin}) shows
again the limitation imposed by the CR constraint on the use of side
information (see Remark \ref{HB}).
\end{rem}
\begin{figure}[h!]
\centering \includegraphics[bb=76bp 350bp 485bp 685bp,clip,scale=0.7]{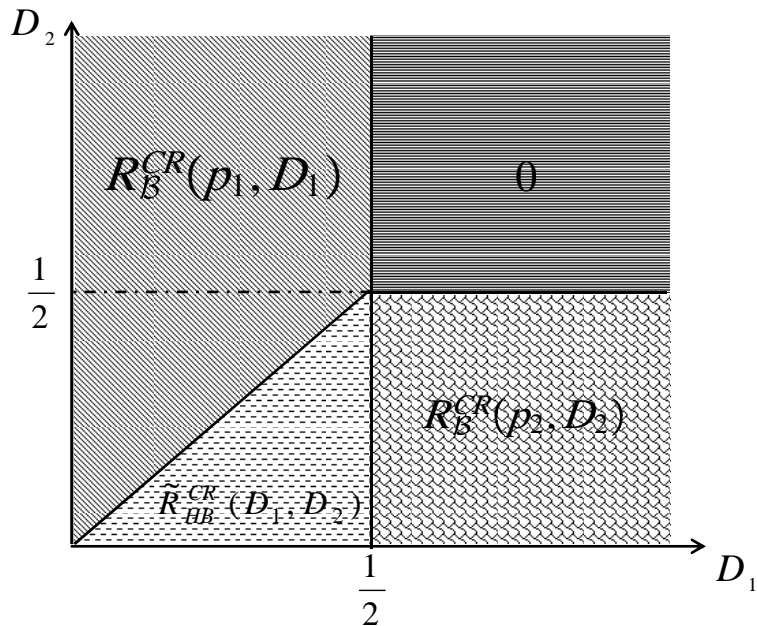}
\caption{Illustration of the distortion regions in the rate-distortion function
(\ref{eq:region-bin}) for a binary source with degraded erased side
information and Hamming distortion.}

\label{fig:region-bin}
\end{figure}

\begin{prop}
\label{prop:HB_Bin}The rate-distortion function for the HB problem
with CR for the binary source with the stochastically degraded erased
side information sequences illustrated in Fig. \ref{fig:test} under
Hamming distortion is given by\textup{
\begin{eqnarray}
R_{HB}^{CR}(D_{1},D_{2}) & = & \left\{ \begin{array}{ll}
0 & \mbox{if }D_{1}\geq1/2\mbox{ and }D_{2}\geq1/2,\\
R_{\mathcal{B}}^{CR}(D_{1},p_{1}) & \mbox{if }D_{1}\leq1/2\mbox{ and }D_{2}\geq\min(D_{1},1/2)\\
R_{\mathcal{B}}^{CR}(D_{2},p_{2}) & \mbox{if \ensuremath{D_{1}\geq1/2}}\mbox{ and }D_{2}\leq1/2\\
\tilde{R}_{HB}^{CR}(D_{1},D_{2}) & \mbox{if }D_{2}\leq D_{1}\leq1/2
\end{array}\right.\label{eq:region-bin}
\end{eqnarray}
}where $R_{\mathcal{B}}^{CR}(D,N)$ is defined in (\ref{eq:RCR_Bin})
and\textup{
\begin{align}
\tilde{R}_{HB}^{CR}(D_{1},D_{2}) & \overset{\bigtriangleup}{=}p_{1}(1-H(D_{1}))+p_{2}(H(D_{1})-H(D_{2})).\label{eq:4_region-bin}
\end{align}
}Moreover, for the same source under erasure distortion the rate-distortion
function is given by (\ref{eq:region-bin}) by substituting $R_{\mathcal{B}}^{CR}(D_{j},p_{j})$
with $R_{\mathcal{BE}}^{CR}(D_{j},p_{j})$ as defined in (\ref{eq:RCR_Bin-1})
for $j=1,2$ and by substituting (\ref{eq:4_region-bin}) with\textup{
\begin{align}
\tilde{R}_{HB,E}^{CR}(D_{1},D_{2}) & \overset{\bigtriangleup}{=}p_{1}(1-D_{1})+p_{2}(D_{1}-D_{2}).
\end{align}
}
\end{prop}
Similar to the Gaussian example, the characterization of the rate
distortion function (\ref{eq:region-bin}) requires different considerations
for the four subregions of the $(D_{1},D_{2})$ plane sketched in
Fig. \ref{fig:region-bin}. In fact, for $D_{1}\geq1/2\mbox{ and }D_{2}\geq1/2$,
the required rate is zero, since the distortion constraints are trivially
met by setting $\hat{X_{1}}=\hat{X_{2}}=0$ in the achievable rate
(\ref{eq:RHB}). For the case $D_{1}\geq1/2\mbox{ and }D_{2}\leq1/2$,
it is sufficient to cater only to Decoder 2 by setting $\hat{X_{1}}=0$
and $X=\hat{X}_{2}\oplus Q_{2}$, with $Q_{2}\sim\textrm{Ber}(D_{2})$
independent of $X$, in the achievable rate (\ref{eq:RHB}). That
this rate cannot be improved upon is a consequence from the trivial
converse
\begin{align}
R_{HB}^{CR}(D_{1},D_{2})\geq\textrm{max}\{R_{\mathcal{B}}^{CR}(D_{1},p_{1}),R_{\mathcal{B}}^{CR}(D_{2},p_{2})\},
\end{align}
which follows by cut-set arguments. The same converse suffices also
for the regime $D_{1}\leq1/2\mbox{ and }D_{2}\geq\min(D_{1},1/2)$.
For this case, achievability follows by setting $X=\hat{X}_{1}\oplus Q_{1}$
and $\hat{X_{1}}=\hat{X_{2}}$ in (\ref{eq:RHB}), where $Q_{1}\sim\textrm{Ber}(D_{1})$
is independent of $\hat{X}_{1}$. In the remaining case, namely $D_{2}\leq D_{1}\leq1/2$,
the rate-distortion function does not follow from the point-to-point
result (\ref{eq:RCR_Bin}) as for the regimes discussed thus far.
The analysis of this case can be found in Appendix D. Similar arguments
apply also for the erasure distortion metric.

We now compare the rate-distortion function for the binary source
$X\sim\textrm{Ber}(\frac{1}{2})$ with erased side information under
Hamming distortion for three settings. In the first setting, known
as the Kaspi model \cite{Kaspi}, the encoder knows the side information,
and thus the position of the erasures. For this case, the rate-distortion
function $R_{Kaspi}(D_{1},D_{2})$ for the example at hand was calculated
in \cite{Perron}. Note that in the Kaspi model, the CR constraint
does not affect the rate-distortion performance since the encoder
has all the information available at the decoders. The second model
of interest is the standard HB setting with no CR constraint, whose
rate-distortion function $R_{HB}(D_{1},D_{2})$ for the example at
hand was derived \cite{Ravi}. The third model is the HB setup with
CR studied here. We clearly have the inequalities
\begin{align}
R_{Kaspi}(D_{1},D_{2})\leq R_{HB}(D_{1},D_{2})\leq R_{HB}^{CR}(D_{1},D_{2}),\label{eq:compare}
\end{align}
where the first inequality in (\ref{eq:compare}) accounts for the
impact of the availability of the side information at the encoder,
while the second reflects the potential performance loss due to the
CR constraint. 

\begin{figure}[h!]
\centering \includegraphics[bb=47bp 280bp 558bp 669bp,clip,scale=0.7]{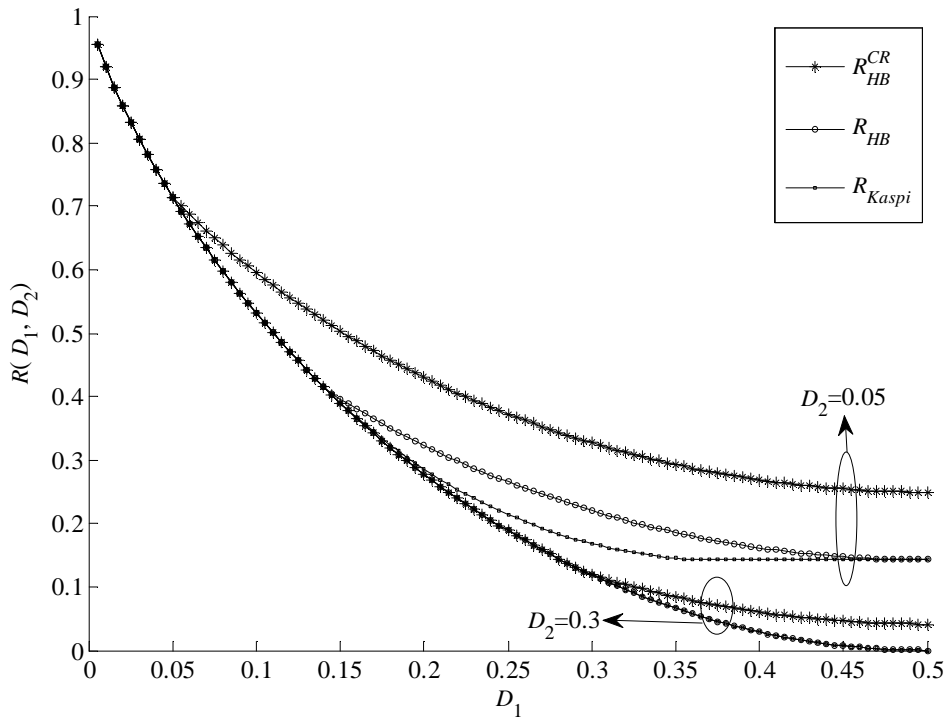}
\caption{Rate-distortion functions $R_{Kaspi}(D_{1},D_{2})$ \cite{Perron},
$R_{HB}(D_{1},D_{2})$ \cite{Ravi} and $R_{HB}^{CR}(D_{1},D_{2})$
(\ref{eq:region-bin}) for a binary source under erased side information
versus distortion $D_{1}$ ($p_{1}=1$, $p_{2}=0.35$, $D_{2}=0.05$
and $D_{2}=0.3$).}

\label{fig:plot3}
\end{figure}

Fig. \ref{fig:plot3} shows the aforementioned rate-distortion functions
with $p_{1}=1$ and $p_{2}=0.35$, which corresponds to the case where
Decoder 1 has no side information, for two values of the distortion
$D_{2}$ versus the distortion $D_{1}$. For $D_{2}\geq\frac{p_{2}}{2}=0.175$,
the given settings reduce to one in which the encoder needs to communicate
information only to Decoder 1. Since Decoder 1 has no side information,
the Kaspi and HB settings yield equal performance i.e., $R_{Kaspi}(D_{1},D_{2})=R_{HB}(D_{1},D_{2})$.
Moreover, if $D_{1}$ is sufficiently smaller than $D_{2}$, the operation
of the encoder is limited by the distortion requirements of Decoder
1. In this case, Decoder 2 can in fact reconstruct as $\hat{X}_{1}=\hat{X}_{2}$
while still satisfying its distortion constraints. Therefore,  we
obtain the same performance in all of the three settings, i.e., $R_{Kaspi}(D_{1},D_{2})=R_{HB}(D_{1},D_{2})=R_{HB}^{CR}(D_{1},D_{2}).$
We also note the general performance loss due to the CR constraint,
unless, as discussed above, distortion $D_{1}$ is sufficiently smaller
than $D_{2}$.

\section{Heegard-Berger Problem with Cooperative Decoders\label{sec:HB-cop}}

The system model for the HB problem with CR and decoder cooperation
is similar to the one provided in Sec. \ref{sub:System-model_HB}
with the following differences. Here, in addition to encoding function
given in (\ref{eq:enc}) which maps the source sequence $X^{n}$ into
a message $\mbox{\ensuremath{J_{1}}}$ of $nR_{1}$ bits, there is
an encoder at Decoder 1 given by 
\begin{align}
g_{1}\textrm{: }[1,2^{nR_{1}}]\times\mathcal{Y}_{1}^{n}\rightarrow[1,2^{nR_{2}}],
\end{align}
which maps message $\ensuremath{J_{1}}$ and the source sequence $Y_{1}^{n}$
into a message $\mbox{\ensuremath{J_{2}}}.$ Moreover, instead of
the decoding function given in (\ref{eq:dec1}), we have the decoding
function for Decoder 2
\begin{align}
h_{2}\textrm{: }[1,2^{nR_{1}}]\times[1,2^{nR_{2}}]\times\mathcal{Y}_{2}^{n}\rightarrow\mathcal{\hat{X}}_{2}^{n},\label{eq:dec1-1}
\end{align}
which maps the messages $J_{1}$ and $J_{2}$ and the side information
$Y_{2}^{n}$ into the estimated sequence $\hat{X}_{2}^{n}$.

\subsection{Rate-Distortion Region for $X-Y_{1}-Y_{2}$\label{sub:xy1y2_cop}}

In this section, a single-letter characterization of the rate-distortion
region is derived under the assumption that the joint pmf $p(x,y_{1},y_{2})$
is such that the Markov chain $X-Y_{1}-Y_{2}$ holds%
\footnote{Note that, unlike the conventional HB problem studied in Sec. \ref{sec:HB},
the rate-distortion region with cooperative decoders depends on the
joint distribution of the variables ($Y_{1},Y_{2}$), and thus stochastic
and physical degradedness of the side information sequences lead to
different results.%
}. 
\begin{prop}
The rate-distortion region $\mathcal{R^{\textrm{\ensuremath{CR}}}\mbox{\ensuremath{(D_{1},D_{2})}}}$
for the HB source coding problem with CR and cooperative decoders
under the assumption $X-Y_{1}-Y_{2}$ is given by the union of all
rate pairs $(R_{1},R_{2})$ that satisfy the conditions\begin{subequations}\label{eqn: reg_cop}
\begin{eqnarray}
R_{1} & \geq & I(X;\hat{X}_{1}\hat{X}_{2}|Y_{1})\label{eq:R1_cop}\\
\mbox{ and }R_{1}+R_{2} & \geq & I(X;\hat{X}_{2}|Y_{2})+I(X;\hat{X}_{1}|Y_{1},\hat{X}_{2}),\label{eq:R1+R2_cop}
\end{eqnarray}
\end{subequations}where the mutual information terms are evaluated
with respect to the joint pmf
\begin{align}
p(x,y_{1},y_{2},\hat{x}_{1},\hat{x}_{2})=p(x,y_{1})p(y_{2}|y_{1})p(\hat{x}_{1},\hat{x}_{2}|x) & ,\label{eq:joint-2-1}
\end{align}
for some pmf $p(\hat{x}_{1},\hat{x}_{2}|x)$ such that the constraints
(\ref{eq:const}) are satisfied.
\end{prop}
The proof of the converse can be easily established following cut-set
arguments for bound (\ref{eq:R1_cop}), while the bound (\ref{eq:R1+R2_cop})
on the sum-rate $R_{1}+R_{2}$ can be proved following the same step
as in Appendix A and substituting $J$ with $(J_{1},J_{2})$. As for
the achievability, it follows as a straightforward extension of \cite[Sec. III]{Vasudevan_cop}
to the setup at hand where Decoder 2 has side information as well.
It is worth emphasizing that the reconstruction $\hat{X}_{2}$ for
the Decoder 2, which has degraded side information, is conveyed by
using both the direct link from the Encoder, of rate $R_{1}$, and
the path Encoder-Decoder 1-Decoder 2. The latter path leverages the
the better side information at Decoder 1 and the cooperative link
of rate $R_{2}$. 
\begin{rem}
If we remove the CR constraint, the problem of determining the rate-distortion
region for the setting of Fig. \ref{fig:fig2_cop} under the assumption
$X-Y_{1}-Y_{2}$ is still open. In \cite{Vasudevan_cop}, inner and
outer bounds are obtained to the rate distortion region, for the case
which the side information $Y_{2}$ is absent. The bounds were shown
to coincide for the case where Decoder 1 wishes to recover $X$ losslessly
(i.e., $D_{1}=0$) and also for certain distortion regimes in the
quadratic Gaussian case. Moreover, the rate distortion tradeoff is
completely characterized in \cite{Vasudevan_cop} for the case in
which the encoder also has access to the side information. We note
that, as per the discussion in Sec. \ref{sub:binary-HB}, these latter
result immediately carry over to the case with CR constraint since
the encoder is informed about the side information. 
\end{rem}
\begin{rem}To understand why imposing the CR constraint simplifies
the problem of obtaining a single-letter characterization of the rate-distortion
function, let us consider the degrees of freedom available at Decoder
1 in Fig. 3 for the use of the link of rate $R_{2}$. In general,
Decoder 1 can follow two possible strategies: the first is forwarding,
whereby Decoder 1 simply forwards some of the bits received from the
encoder to Decoder 2; while the second is recompression, whereby the
data received from the encoder is combined with the available side
information $Y{}_{1}^{n}$, compressed to at most $R_{2}$ bits per
symbol, and then sent to Decoder 2. It is the interplay and contrast
between these two strategies that makes the general problem hard to
solve. In particular, while the strategies of forwarding/recompression
and combinations thereof appear to be natural candidates for the problem,
finding a matching converse when both such degrees of freedom are
permissible at the decoder is difficult (see, e.g., \cite{Cuff}).
However, under the CR constraint, the strategy of recompression becomes
irrelevant, since any information about the side information $Y{}_{1}^{n}$
that is not also available at the encoder cannot be leveraged by Decoder
2 without violating the CR constraint. This restriction in the set
of available strategies for Decoder 1 makes the problem easier to
address under the CR constraint.\textquotedblright{}

\end{rem}

\subsection{Rate-Distortion Region for $X-Y_{2}-Y_{1}$\label{sub:xy2y1_cop}}

In this section, a single-letter characterization of the rate-distortion
region is derived under the assumption that the joint pmf $p(x,y_{1},y_{2})$
is such that the Markov chain relationship $X-Y_{2}-Y_{1}$ holds. 
\begin{prop}
The rate-distortion region $\mathcal{R^{\textrm{\ensuremath{CR}}}\mbox{\ensuremath{(D_{1},D_{2})}}}$
for the HB source coding problem with CR and cooperative decoders
under the assumption the Markov chain relationship $X-Y_{2}-Y_{1}$
is given by the union of all rate pairs $(R_{1},R_{2})$ that satisfy
the conditions\begin{subequations}\label{eqn: reg_cop-1}
\begin{eqnarray}
R_{1} & \geq & I(X;\hat{X}_{1}|Y_{1})+I(X;\hat{X}_{2}|Y_{2},\hat{X}_{1})\label{eq:R1_cop-1}\\
\mbox{ and }R_{2} & \geq & 0,\label{eq:R2_cop1}
\end{eqnarray}
\end{subequations}where the mutual information terms are evaluated
with respect to the joint pmf
\begin{align}
p(x,y_{1},y_{2},\hat{x}_{1},\hat{x}_{2})=p(x,y_{2})p(y_{1}|y_{2})p(\hat{x}_{1},\hat{x}_{2}|x) & ,\label{eq:joint-2-1-1}
\end{align}
for some pmf $p(\hat{x}_{1},\hat{x}_{2}|x)$ such that the constraints
(\ref{eq:const}) are satisfied.
\end{prop}
The proof of achievability follows immediately by neglecting the link
of rate $R_{2}$ and using rate $R_{1}$ as per the HB scheme of Proposition
\ref{prop:HB_converse}. The converse follows by considering an enhanced
system in which Decoder 2 is provided with the side information of
Decoder 1. In this system, link $R_{2}$ becomes useless since Decoder
2 possesses all the information available at Decoder 1. It follows
that the system reduces to the HB problem with degraded sources studied
in the previous section and the bound (\ref{eq:R1_cop-1}) follows
immediately from Proposition \ref{prop:HB_converse}.
\begin{rem}
In the case without CR, the rate-distortion function is given similarly
to (\ref{eqn: reg_cop-1}), but with the HB rate-distortion function
(\ref{eq:RHB-1}) in lieu of the rate-distortion function of the HB
problem with CR in (\ref{eq:R1_cop-1}). 
\end{rem}

\section{Cascade Source Coding with Common Reconstruction \label{sec:Cascade-Source-Coding}}

In this section, we first detail the system model in Fig. \ref{fig:cascade}
of cascade source coding with CR. As mentioned in Sec. I, the motivation
for studying this class of models comes from multi-hop applications.
Next, the characterization of the corresponding rate-distortion performance
is presented under the assumption that one of the two side information
sequences is a degraded version of the other. Finally, following the
previous section, three specific examples are worked out, namely Gaussian
sources under quadratic distortion (Sec. \ref{sub:Gaussian-cascade}),
and binary sources with side information subject to erasures under
Hamming or erasure distortion (Sec. \ref{sub:binary-cascade}).

\subsection{System model}

In this section, the system model for the cascade source coding problem
with CR is detailed similar to Sec. \ref{sub:System-model_HB}. The
problem is defined by the pmf $p_{XY_{1}Y_{2}}(x,y_{1},y_{2})$ and
discrete alphabets $\mathcal{X},\mathcal{Y}_{1},\mathcal{Y}_{2},\hat{\mathcal{X}}_{1},$
and $\hat{\mathcal{X}}_{2}$ as follows. The source sequence $X^{n}$
and side information sequences $Y_{1}^{n}$ and $Y_{2}^{n}$, with
$X^{n}\in\mathcal{X}^{n}$, $Y_{1}^{n}\in\mathcal{Y}_{1}^{n}$, and
$Y_{2}^{n}\in\mathcal{Y}_{2}^{n}$ are such that the tuples $(X_{i},Y_{1i},Y_{2i})$
for $i\in[1,n]$ are i.i.d. with joint pmf $p_{XY_{1}Y_{2}}(x,y_{1},y_{2})$.
Node 1 measures a sequence $X^{n}$ and encodes it into a message
$J_{1}$ of $nR_{1}$ bits, which is delivered to Node 2. Node 2 estimates
a sequence $\hat{X}_{1}^{n}\in\mathcal{\hat{X}}_{1}^{n}$ within given
distortion requirements. Node 2 also encodes the message $J_{1}$
received from Node 1 and the sequence $Y_{1}^{n}$ into a message
$J_{2}$ of $nR_{2}$ bits, which is delivered to Node 3. Node 3 estimates
a sequence $\hat{X}_{2}^{n}\in\mathcal{\hat{X}}_{2}^{n}$ within given
distortion requirements. Distortion and CR requirements are defined
as in Sec. \ref{sub:System-model_HB}, leading to the following definition. 
\begin{defn}
\label{cascade_code_def}An $(n,R_{1},R_{2},D_{1},D_{2},\epsilon)$
code for the cascade source coding problem with CR consists an encoding
function for Node 1,
\begin{align}
g_{1}\textrm{: }\mathcal{X}^{n}\rightarrow[1,2^{nR_{1}}],
\end{align}
which maps the source sequence $X^{n}$ into a message $\mbox{\ensuremath{J_{1}};}$
an encoding function for Node 2, 
\begin{align}
g_{2}\textrm{: }[1,2^{nR_{1}}]\times\mathcal{Y}_{1}^{n}\rightarrow[1,2^{nR_{2}}],
\end{align}
which maps the source sequence $Y_{1}^{n}$ and message $J_{1}$ into
a message $\mbox{\ensuremath{J_{2}};}$ a decoding function for Node
2
\begin{align}
h_{1}\textrm{: }[1,2^{nR_{1}}]\times\mathcal{Y}_{1}^{n}\rightarrow\mathcal{\hat{X}}_{1}^{n},
\end{align}
which maps message $J_{1}$ and the side information $Y_{1}^{n}$
into the estimated sequence $\hat{X}_{1}^{n}$; a decoding function
for Node 3
\begin{align}
h_{2}\textrm{: }[1,2^{nR_{2}}]\times\mathcal{Y}_{2}^{n} & \rightarrow\mathcal{\hat{X}}_{2}^{n},
\end{align}
which maps message $J_{2}$ and the side information $Y_{2}^{n}$
into the estimated sequence $\hat{X}_{2}^{n}$; two encoder reconstruction
functions as in (\ref{eqn: en_recons}), which map the source sequence
into estimated sequences $\psi_{1}(X^{n})$ and $\psi_{2}(X^{n})$
at Node 1; such that the distortion constraints (\ref{eq:dist_const})
and (\ref{eq:CR_req}) are satisfied.

Given a distortion pair $(D_{1},D_{2})$, a rate pair $(R_{1},R_{2})$
is said to be achievable if, for any $\epsilon>0$ and sufficiently
large $n$, there a exists an $(n,R_{1},R_{2},D_{1}+\epsilon,D_{2}+\epsilon,\epsilon)$
code. The rate-distortion region $\mathcal{R}(D_{1},D_{2})$ is defined
as the closure of all rate pairs $(R_{1},R_{2})$ that are achievable
given the distortion pair $(D_{1},D_{2})$.
\end{defn}

\subsection{Rate-Distortion Region for $X-Y_{1}-Y_{2}$\label{sub:xy1y2}}

In this section, a single-letter characterization of the rate-distortion
region is derived under the assumption that the joint pmf $p(x,y_{1},y_{2})$
is such that the Markov chain relationship $X-Y_{1}-Y_{2}$ holds
\footnote{As for the HB problem with cooperative decoders studied in Sec. \ref{sec:HB-cop},
the rate-distortion region of the cascade source coding problem depends
on the joint distribution of the variables ($Y_{1},Y_{2}$), and thus
stochastic and physical degradedness of the side information sequences
lead to different results.%
}. 
\begin{prop}
The rate-distortion region $\mathcal{R^{\textrm{\ensuremath{CR}}}\mbox{\ensuremath{(D_{1},D_{2})}}}$
for the cascade source coding problem with CR is given by the union
of all rate pairs $(R_{1},R_{2})$ that satisfy the conditions\begin{subequations}\label{eqn: outer-1}
\begin{eqnarray}
R_{1} & \geq & I(X;\hat{X}_{1}\hat{X}_{2}|Y_{1})\label{eq:ach_R1-1}\\
\mbox{ and }R_{2} & \geq & I(X;\hat{X}_{2}|Y_{2}),\label{eq:ach_R2-1}
\end{eqnarray}
\end{subequations}where the mutual information terms are evaluated
with respect to the joint pmf
\begin{align}
p(x,y_{1},y_{2},\hat{x}_{1},\hat{x}_{2})=p(x,y_{1})p(y_{2}|y_{1})p(\hat{x}_{1},\hat{x}_{2}|x) & ,\label{eq:joint-2}
\end{align}
for some pmf $p(\hat{x}_{1},\hat{x}_{2}|x)$ such that the constraints
(\ref{eq:const}) are satisfied.
\end{prop}
The proof of the converse is easily established following cut-set
arguments. To prove achievability, it is sufficient to consider a
scheme based on binning at Node 1 and decode and rebin at Node 2 (see
\cite{Chia}). Specifically, Node 1 randomly generates a standard
lossy source code $\hat{X}_{1}^{n}$ for the source $X^{n}$ with
rate $I(X;\hat{X}_{1})$ bits per source symbol. Random binning is
used to reduce the rate to $I(X;\hat{X}_{1}|Y_{1}).$ Node 1 then
maps the source $X^{n}$ into the reconstruction sequence $\hat{X}_{2}^{n}$
using a codebook that is generated conditional on $\hat{X}_{1}^{n}$
with rate $I(X;\hat{X}_{2}|\hat{X}_{1})$ bits per source symbol.
Using the side information $Y_{1}^{n}$ available at Node 2, random
binning is again used to reduce the rate to $I(X;\hat{X}_{2}|Y_{1}\hat{X}_{1})$.
The codebook of $\hat{X}{}_{2}^{n}$ is also randomly binned to the
rate $I(X;\hat{X}_{2}|Y_{2})$. Node 2, having recovered $\hat{X}_{2}^{n}$,
forwards the corresponding bin index to Node 3. The latter, by choice
of the binning rate, is able to obtain $\hat{X}_{2}^{n}$. Note that,
since the reconstruction sequences $\hat{X}_{1}^{n}$ and $\hat{X}_{2}^{n}$
are generated by the encoder, functions $\psi_{1}$ and $\psi_{2}$
that guarantees the CR constraints (\ref{eq:CR_req}) exist by construction.
\begin{rem}
If we remove the CR constraint, the problem of determining the rate-distortion
region for the setting of Fig. \ref{fig:cascade} under the Markov
condition $X-Y_{1}-Y_{2}$ is still open. In the special case in which
$Y_{1}=Y_{2}$ the problem has been solved in \cite{Vasudevan} for
Gaussian sources under quadratic distortion and in \cite{Ravi} for
binary sources with erased side information under Hamming distortion. 

\begin{rem}\label{causal}Following Remark \ref{causal-0}, if both
side information sequences are causal, it can be shown that they have
no impact on the rate-distortion function (\ref{eqn: outer-1}). Therefore,
the rate-distortion region follows immediately from the results in
(\ref{eqn: outer-1}) by removing both of the side information terms.
Note that with causal side information sequences the rate-distortion
function holds for any joint pmf $p(x,y_{1},y_{2})$ with no degradedness
requirements. Moreover, if only the side information $Y_{2}$ is causal,
while $Y_{1}$ is still observed non-causally, then the side information
$Y_{2}$ can be neglected without loss of optimality, and the rate-distortion
region follows from (\ref{eqn: outer-1}) by removing the conditioning
on $Y_{2}$.

\end{rem}
\end{rem}

\subsection{Bounds on the Rate-Distortion Region for $X-Y_{2}-Y_{1}$\label{sub:xy2y1}}

In this section, outer and inner bounds are derived for the rate-distortion
region under the assumption that the joint pmf $p(x,y_{1},y_{2})$
is such that the Markov chain relationship $X-Y_{2}-Y_{1}$ holds.
The bounds are then shown to coincide in Sec. \ref{sub:Gaussian-cascade}
for Gaussian sources and in Sec. \ref{sub:binary-cascade} for binary
sources with erased side information.

\begin{prop}\label{prop:outer}(Outer bound) The rate-distortion
region $\mathcal{R^{\textrm{\ensuremath{CR}}}\mbox{\ensuremath{(D_{1},D_{2})}}}$
for the cascade source coding problem with CR is contained in the
region $\mathcal{R}_{out}^{\textrm{\ensuremath{CR}}}\mbox{\ensuremath{(D_{1},D_{2})},}$
which is given by the set of all rate pairs $(R_{1},R_{2})$ that
satisfy the conditions \begin{subequations}\label{eqn: outer}
\begin{eqnarray}
R_{1} & \geq & R_{HB}^{CR}(D_{1},D{}_{2})\label{eq:converse_R1}\\
\textrm{and }R_{2} & \geq & R_{X|Y_{2}}^{CR}(D{}_{2}),\label{eq:converse_R2}
\end{eqnarray}
\end{subequations}where $R_{HB}^{CR}(D_{1},D{}_{2})$ is defined
in (\ref{eq:RHB}) and we have $R_{X|Y_{2}}^{CR}(D{}_{2})=\mbox{ min }I(X;\hat{X}_{2}|Y_{2}),$
where the minimization is performed with respect to the conditional
pmf $p(\hat{x}_{2}|x)$ under the distortion constraints (\ref{eq:const})
for $j=2.$

\end{prop}
\begin{prop}
\label{prop:inner}(Inner bound) The rate-distortion region \textup{$\mathcal{R}^{\textrm{\ensuremath{CR}}}\mbox{\ensuremath{(D_{1},D_{2})}}$}
for the cascade source coding problem with CR contains the region
\textup{$\mathcal{R}_{in}^{\textrm{\ensuremath{CR}}}\mbox{\ensuremath{(D_{1},D_{2})},}$}
which is given by the union of all rate pairs $(R_{1},R_{2})$ that
satisfy the conditions \begin{subequations}\label{eqn: inner}
\begin{eqnarray}
R_{1} & \geq & I(X;\hat{X}_{1}|Y_{1})+I(X;\hat{X}_{2}|Y_{2}\hat{X}_{1})\label{eq:ach_R1}\\
\textrm{and }R_{2} & \geq & I(X;\hat{X}_{1}|Y_{2})+I(X;\hat{X}_{2}|\hat{X}_{1}Y_{2})\label{eq:ach_R2}\\
 & = & I(X;\hat{X}_{1}\hat{X}_{2}|Y_{2})
\end{eqnarray}
\end{subequations} where the mutual information terms are evaluated
with respect to the joint pmf
\begin{align}
p(x,y_{1},y_{2},\hat{x}_{1},\hat{x}_{2})=p(x,y_{2})p(y_{1}|y_{2})p(\hat{x}_{1},\hat{x}_{2}|x) & ,\label{eq:joint-3}
\end{align}
for some pmf $p(\hat{x}_{1},\hat{x}_{2}|x_{1})$ such that the distortion
constraints (\ref{eq:const}) are satisfied.
\end{prop}
The outer bound in Proposition \ref{prop:outer} follows immediately
from cut-set arguments similar to those in \cite{Vasudevan} and \cite{Ravi}.
As for the inner bound of Proposition 19, the strategy works as follows.
Node 1 sends the description $\hat{X}_{1}^{n}$ to Node 2 using binning
with rate $I(X;\hat{X}_{1}|Y_{1})$. It also maps the sequence $X^{n}$
into the sequence $\hat{X}_{2}^{n}$ using a conditional codebook
with respect to $\hat{X}_{1}^{n}$, which is binned in order to leverage
the side information $Y_{2}^{n}$ at Node 3 with rate $I(X;\hat{X}_{2}|\hat{X}_{1},Y_{2})$.
Node 2 recovers $\hat{X}_{1}^{n}$, whose codebook is then binned
to rate $I(X;\hat{X}_{1}|Y_{2})$. Then, it forwards the so obtained
bin index for $\hat{X}_{1}^{n}$ and the bin index for the codebook
of $\hat{X}_{2}^{n}$ produced by Node 1 to Node 3. By the choice
of the rates, the latter can recover both $\hat{X}_{1}^{n}$ and $\hat{X}_{2}^{n}$
. Since both descriptions are produced by Node 1, the CR constraint
is automatically satisfied.

The inner and outer bounds defined above do not coincide in general.
However, in the next sections, we provide two examples in which they
coincide and thus characterize the rate-distortion region of the corresponding
settings.
\begin{rem}
Without the CR constraint, the problem of deriving the rate-distortion
region for the setting at hand under the Markov chain condition $X-Y_{2}-Y_{1}$
is open. The problem has been solved in \cite{Vasudevan} for Gaussian
sources under quadratic distortion and in \cite{Ravi} for binary
sources with erased side information under Hamming distortion for
$Y_{1}=Y_{2}$.
\end{rem}

\subsubsection{Gaussian Sources and Quadratic Distortion\label{sub:Gaussian-cascade}}

In this section, we assume the Gaussian sources in (\ref{eqn:side_Gauss})
and the quadratic distortion as in Sec \ref{sub:Gaussian-HB}, and
derive the rate-distortion region for the cascade source coding problem
with CR.
\begin{prop}
\label{prop:cascade_Gauss}The rate-distortion region \textup{$\mathcal{R}^{\textrm{\ensuremath{CR}}}\mbox{\ensuremath{(D_{1},D_{2})}}$}
for the cascade source coding problem with CR for the Gaussian sources
in (\ref{eqn:side_Gauss}) and quadratic distortion is given by (\ref{eqn: outer})
with\textup{ $R_{HB}^{\textrm{\ensuremath{CR}}}\mbox{\ensuremath{(D_{1},D_{2})}}$
}in\textup{ (\ref{eq:region}) }and\textup{ $R_{X|Y_{2}}^{CR}(D{}_{2})=R_{\mathcal{G}}^{\textrm{\ensuremath{CR}}}\mbox{\ensuremath{(D_{2},N_{2})}}$
(}see\textup{ (\ref{eq:RCR_Gauss})).}
\end{prop}
The proof is given in Appendix E.

\subsubsection{Binary Sources with Erased Side Information and Hamming Distortion\label{sub:binary-cascade}}

In this section, we assume the binary sources in Fig. \ref{fig:test}
and the Hamming distortion as in Sec \ref{sub:binary-HB}, and derive
the rate-distortion region for the cascade source coding problem with
CR.
\begin{prop}
\label{prop:cascade_bin}The rate-distortion region \textup{$\mathcal{R}^{\textrm{\ensuremath{CR}}}\mbox{\ensuremath{(D_{1},D_{2})}}$}
for the cascade source coding problem with CR for the binary sources
in Fig. \ref{fig:test} and Hamming distortion is given by (\ref{eqn: outer})
with \textup{$R_{HB}^{CR}(D_{1},D{}_{2})$ in (\ref{eq:region-bin})
and $R_{X|Y_{2}}^{CR}(D{}_{2})=R_{\mathcal{B}}^{\textrm{\ensuremath{CR}}}(D_{2},p_{2})$
(see (\ref{eq:RCR_Bin})).}
\end{prop}
The proof is given in Appendix F.

\section{Heegard-Berger Problem with Constrained Reconstruction\label{sec:ConR}}

In this section, we revisit the HB problem and relax the CR constraint
to the ConR constraint of {[}13{]}. This implies that we still adopt
the code as per Definition 1, but we substitute (\ref{eq:CR_req})
with the less stringent constraint
\begin{align}
\frac{1}{n}\sum_{i=1}^{n}\textrm{E}\left[d_{e,j}(\hat{X}_{ji},\psi_{ji}(X^{n}))\right] & \leq D_{e,j}\mbox{ for \ensuremath{j=1,2}},\label{eq:dist_const_ext}
\end{align}
where $d_{e,j}(\hat{x}_{j},\hat{x}_{e,j})\textrm{: }\hat{\mathcal{X}}_{j}\times\hat{\mathcal{X}}_{j}\rightarrow[0,D_{e,max}]$
is a per-symbol distortion metric and we have used $\psi_{ji}(X^{n})$,$\mbox{ for \ensuremath{j=1,2}},$
to denote the $i$th letter of the vector $\psi_{j}(X^{n})=(\psi_{j1}(X^{n}),...,\psi_{jn}(X^{n}))$. 
\begin{defn}
Given a distortion tuple $(D_{e,1},D_{e,2},D_{1},D_{2})$, a rate
$R$ is said to be achievable if, for any $\epsilon>0$ and sufficiently
large $n$, there a exists an $(n,R,D_{e,1}+\epsilon,D_{e,2}+\epsilon,D_{1}+\epsilon,D_{2}+\epsilon,\epsilon)$
code. The rate-distortion function $R(D_{e,1},D_{e,2},D_{1},D_{2})$
is defined as $R(D_{e,1},D_{e,2},D_{1},D_{2})=$ inf$\{R\textrm{:}$
the tuple $(D_{e,1},D_{e,2},D_{1},D_{2})$ is achievable\}.
\end{defn}
Note that, by setting $D_{e,j}=0\mbox{ for \ensuremath{j=1,2}},$
and letting $d_{e,j}(\hat{x}_{j},\hat{x}_{e,j})$ be the Hamming distortion
metric (i.e., $d_{e,j}(\hat{x}_{j},\hat{x}_{e,j})=1$ if $x\neq\hat{x}_{j}$
and $d_{e,j}(\hat{x}_{j},\hat{x}_{e,j})=0$ if $x=\hat{x}_{j}$),
we obtain a relaxed CR constraint in which the average per-symbol,
rather than per-block, error probability criterion is adopted. 
\begin{rem}
The problem at hand reduces to the one studied in \cite{Lapidoth}
by setting $D_{1}=D_{max}$ and $D_{e,1}=D_{e,max}$.\end{rem}
\begin{prop}
\label{prop:HB_converse_ext}If the side information $Y_{1}$ is stochastically
degraded with respect to $Y_{2}$, the rate-distortion function for
the HB problem with ConR is given by \begin{subequations}\label{eqn: 2RHB_ext}
\begin{eqnarray}
R_{HB}^{ConR}(D_{e,1},D_{e,2},D_{1},D_{2}) & = & \min I(X;U_{1}|Y_{1})+I(X;U_{2}|Y_{2}U_{1})\label{eq:RHB_ext}\\
 & = & \min I(X;U_{1}U_{2}|Y_{2})+I(U_{1};Y_{2}|Y_{1}),\label{eq:RHB_ext1}
\end{eqnarray}
\end{subequations}where the mutual information terms are evaluated
with respect to the joint pmf
\begin{align}
p(x,y_{1},y_{2},u_{1},u_{2})=p(x,y_{1},y_{2})p(u_{1},u_{2}|x),\label{eq:joint-4}
\end{align}
and minimization is performed with respect to the conditional pmf
$p(u_{1},u_{2}|x)$ and the deterministic functions \textup{$\mathrm{\mathrm{\hat{x}_{j}}\mbox{\ensuremath{(u_{j},y_{j})\textrm{: }\mathcal{U}_{j}\times\mathcal{Y}_{j}\rightarrow\mathcal{\hat{X}}_{j}}}}$
and $\mathrm{\mathrm{\hat{x}_{e,j}}\mbox{\ensuremath{(u_{j},x)\textrm{: }\mathcal{U}_{j}\times\mathcal{X}\rightarrow\mathcal{\hat{X}}_{e,j}}}}$}
for $j=1,2$, such that the distortion constraints\textup{ $\textrm{E}[d_{j}(X,\hat{\textrm{x}}_{j}(U_{j},Y_{j}))]\leq D_{j}$
for $j=1,2$,} and the ConR requirements \textup{
\begin{align}
\mathrm{E}[d_{e,j}(\hat{\textrm{x}}_{j}(U_{j},Y_{j}),\hat{\textrm{x}}_{e,j}(U_{j},X))]\leq D_{e,j},\mbox{ for }j=1,2,\label{eq:const_ext}
\end{align}
}are satisfied. Finally, $(U_{1},U_{2})$ are auxiliary random variables
whose alphabet cardinalities can be constrained as $|\mathcal{U}_{1}|\leq|\mathcal{X}|+4$
and $|\mathcal{U}_{2}|\leq(|\mathcal{X}|+2)^{2}$. 
\end{prop}
The proof is given in Appendix G.
\begin{rem}
Proposition \ref{prop:HB_converse_ext} reduces to \cite[Theorem 2]{Lapidoth}
when setting $D_{1}=D_{max}$ and $D_{e,1}=D_{e,max}$.

\begin{rem}\label{rem20}Similar to \cite[Theorem 2]{Lapidoth},
it can be proved that, by setting $D_{e,1}=D_{e,2}=0$ and letting
$d_{e,j}$ be the Hamming distortion for $j=1,2$, the rate-distortion
function (\ref{eqn: 2RHB_ext}), $R_{HB}^{ConR}(0,0,D_{1},D_{2})$,
reduces to the rate-distortion function with CR (\ref{eq:RHB}). 
\end{rem}
\end{rem}
\begin{rem}
Similar to Remark \ref{rem20}, if $D_{e,1}=0$ and $D_{e,2}=D_{e,max}$,
the rate-distortion function (\ref{eqn: 2RHB_ext}) is given by 
\begin{eqnarray}
R_{HB}^{CR}(0,D_{e,max},D_{1},D_{2}) & = & \min I(X;\hat{X}_{1}|Y_{1})+I(X;U_{2}|Y_{2}\hat{X}_{1}),\label{eq:RHB_ext-1}
\end{eqnarray}
where the mutual information terms are evaluated with respect to the
joint pmf
\begin{align}
p(x,y_{1},y_{2},u_{2},\hat{x}_{1})=p(x,y_{1},y_{2})p(\hat{x}_{1},u_{2}|x),\label{eq:joint-4-1}
\end{align}
and minimization is performed with respect to the conditional pmf
$p(\hat{x}_{1},u_{2}|x)$ and the deterministic functions $\mathrm{\mathrm{\hat{x}_{2}}\mbox{\ensuremath{(u_{2},y_{2})\textrm{: }\mathcal{U}_{2}\times\mathcal{Y}_{2}\rightarrow\mathcal{\hat{X}}_{2}}}}$
and $\mathrm{\mathrm{\hat{x}_{e,2}}\mbox{\ensuremath{(u_{2},x)\textrm{: }\mathcal{U}_{2}\times\mathcal{X}\rightarrow\mathcal{\hat{X}}_{e,2}}}}$,
such that the distortion constraints $\textrm{E}[d_{1}(X,\hat{X}_{1})]\leq D_{1}$
and $\textrm{E}[d_{2}(X,\hat{\textrm{x}}_{2}(U_{2},Y_{2}))]\leq D_{2}$
and the ConR requirement $\mathrm{E}[d_{e,2}(\hat{\textrm{x}}_{2}(U_{2},Y_{2}),\hat{x}_{e,2}(U_{2},X))]\leq D_{e,2}$
are satisfied. It can be proved that this is also the rate-distortion
function under the \textit{partial CR} requirement that there exists
a function $\psi_{1}(X^{n})$ such that (\ref{eq:CR_req}) holds for
$j=1$ only. Similar conclusions apply symmetrically to the case where
CR and ConR requirements are imposed only on the reconstruction of
Decoder 2.

\begin{rem}\label{causal_ext}If both side information sequences
are causally available at the decoders, it can be proved that they
have no impact on the rate-distortion function (\ref{eqn: 2RHB_ext}).
In this case, the rate-distortion function follows immediately from
the results in (\ref{eqn: 2RHB_ext}) by removing conditioning on
both side information sequences. Moreover, the result can be simplified
by introducing a single auxiliary random variable. Similarly, if only
side information $Y_{1}$ is causal, then it can be neglected with
no loss of optimality, and the results follow from (\ref{eqn: 2RHB_ext})
by removing the conditioning on $Y_{1}$.

\end{rem}
\end{rem}
\begin{rem}We note that the ConR formulation studied in this section
is more general than the conventional formulation with distortion
constraints for the decoders only. Therefore, problems that are open
with the conventional formulation, such as HB with cooperative decoders
(Sec. \ref{sec:HB-cop}) and cascade source coding (Sec. \ref{sec:Cascade-Source-Coding}),
are \emph{a fortiori} also open in the ConR set-up.\end{rem}

\section{Concluding Remarks}

The Common Reconstruction requirement \cite{Steinberg}, and its generalization
in \cite{Lapidoth}, substantially modify the problem of source coding
in the presence of side information at the decoders. From a practical
standpoint, in various applications, such as transmission of medical
records, CR is a design constraint. In these cases, evaluation of
the rate-distortion performance under CR thus reveals the cost, in
terms of transmission resources, associated with this additional requirement.
From a theoretical perspective, adding the CR constraint to standard
source coding problems with decoder side information proves instrumental
in concluding about the optimality of various known strategies in
settings in which the more general problem, without the CR constraint,
is open \cite{Steinberg}. This paper has extended these considerations
from a point-to-point setting to three baseline multiterminal settings,
namely the Heegard-Berger problem, the HB problem with cooperating
decoders and the cascade problems. The optimal rate-distortion trade-off
has been derived in a number of cases and explicitly evaluated in
various examples. 

A general subject of theoretical interest is identifying those models
for which the CR requirements enables a solution of problems that
have otherwise resisted solutions for decades. Examples include the
Heegard-Berger and cascade source coding problems with no assumptions
on side information degradedness and the one-helper lossy source coding
problem.

\appendices{ }

\section*{Appendix A: Proof of Proposition \ref{prop:HB_converse}}

We first observe that from Definition \ref{HBcode_def}, since distortion
and CR constraints (\ref{eq:dist_const}) and (\ref{eq:CR_req}) depend
only on the marginal pmfs $p(x,y_{1})$ and $p(x,y_{2}),$ so does
the rate-distortion function. Therefore, in the proof, we can assume,
without loss of generality, that the joint pmf $p(x,y_{1},y_{2})$
satisfies the Markov chain condition $X-Y_{2}-Y_{1}$ so that it factorizes
as (cf. (\ref{eq:degraded}))
\begin{align}
p(x,y_{1},y_{2}) & =p(x,y_{2})\tilde{p}(y_{1}|y_{2}).\label{eq:degraded_joint}
\end{align}

Consider an $(n,R,D_{1}+\epsilon,D_{2}+\epsilon,\epsilon)$ code,
whose existence is required for achievability by Definition \ref{HBcode_def}.
By the CR requirements (\ref{eq:CR_req}), we first observe that we
have the Fano inequalities 
\begin{equation}
H(\psi_{j}(X^{n})|h_{j}(g(X^{n}),Y_{j}^{n}))\leq n\delta(\epsilon),\textrm{ for }j=1,2,\label{eq:Fano}
\end{equation}
for $n$ sufficiently large, where $\delta(\epsilon)=n\epsilon\mathscr{\textrm{log}}|\mathcal{X}|+H_{b}(\epsilon)$.
Moreover, we can write\begin{subequations}
\begin{align}
nR & =H(J)\geq H(J|Y_{1}^{n})\\
 & \overset{(a)}{=}H(J|Y_{1}^{n}Y_{2}^{n})+I(J;Y_{2}^{n}|Y_{1}^{n}),\label{eq:H(J)}
\end{align}
\end{subequations}where ($a$) follows by the definition of mutual
information. From now on, to simplify notation, we do not make explicit
the dependence of $\psi_{j}$, $g_{j}$ and $h_{j}$ on $X^{n}$ and
$(J,Y_{j}^{n})$, respectively. We also define $\psi_{ji}$ as the
$i$th symbol of the sequence $\psi_{j}$ so that $\psi_{j}=(\psi_{j1},...,\psi_{jn})$.

The first term in (\ref{eq:H(J)}), $H(J|Y_{1}^{n}Y_{2}^{n})$, can
be treated as in {\cite[Sec. V.A.]{Steinberg}}, or, more simply,
we can proceed as follows:\begin{subequations}
\begin{eqnarray}
H(J|Y_{1}^{n}Y_{2}^{n}) & \overset{(a)}{=} & I(J;X^{n}|Y_{1}^{n}Y_{2}^{n})\\
 & \overset{(b)}{\geq} & I(h_{1}h_{2};X^{n}|Y_{1}^{n}Y_{2}^{n})\\
 & = & I(h_{1}h_{2}\psi_{1}\psi_{2};X^{n}|Y_{1}^{n}Y_{2}^{n})-I(\psi_{1}\psi_{2};X^{n}|Y_{1}^{n}Y_{2}^{n}h_{1}h_{2})\\
 & \overset{(c)}{\geq} & I(\psi_{1}\psi_{2};X^{n}|Y_{1}^{n}Y_{2}^{n})-I(\psi_{1}\psi_{2};X^{n}|Y_{1}^{n}Y_{2}^{n}h_{1}h_{2})\\
 & \overset{(d)}{=} & I(\psi_{1}\psi_{2};X^{n}|Y_{2}^{n})-H(\psi_{1}\psi_{2}|Y_{1}^{n}Y_{2}^{n}h_{1}h_{2})+H(\psi_{1}\psi_{2}|Y_{1}^{n}Y_{2}^{n}h_{1}h_{2}X^{n})\\
 & \overset{(e)}{\geq} & I(\psi_{1}\psi_{2};X^{n}|Y_{2}^{n})-n\delta(\epsilon)\\
 & \overset{(f)}{\geq} & \sum_{i=1}^{n}I(\psi_{1i}\psi_{2i};X_{i}|Y_{2i})-n\delta(\epsilon),\label{eq:H(J)1}
\end{eqnarray}
\end{subequations}where ($\mathit{a}$) follows because $J$ is a
function of $X$$^{n}$; ($b$) follows since $h_{1}$ and $h_{2}$
are functions of $(J,Y_{1}^{n})$ and $(J,Y_{2}^{n})$, respectively
; ($c$) follows by using the Markov chain\emph{ $(\psi_{1},\psi_{1},X^{n})\textrm{---}Y_{2}^{n}\textrm{---}Y_{1}^{n}$};
($d$) follows by the chain rule of mutual information and since mutual
information is non-negative; ($e$) follows by (\ref{eq:Fano}) and
since entropy is non-negative; and ($f$) follows by the chain rule
for entropy, since $X^{n}$ and $Y_{2}^{n}$ are i.i.d., and due to
the fact conditioning decreases entropy.

Similarly, the second term in (\ref{eq:H(J)}), namely, $I(J;Y_{2}^{n}|Y_{1}^{n})$,
leads to \begin{subequations}
\begin{align}
I(J;Y_{2}^{n}|Y_{1}^{n}) & \overset{(a)}{\geq}I(\ce{h}_{1};Y_{2}^{n}|Y_{1}^{n})\\
 & =I(\ce{h}_{1}\psi_{1};Y_{2}^{n}|Y_{1}^{n})-I(\psi_{1};Y_{2}^{n}|Y_{1}^{n}h_{1})\\
 & \overset{(b)}{\geq}I(\psi_{1};Y_{2}^{n}|Y_{1}^{n})-H(\psi_{1}|Y_{1}^{n}h_{1})+H(\psi_{1}|Y_{1}^{n}Y_{2}^{n}h_{1})\\
 & \overset{(c)}{\geq}I(\psi_{1};Y_{2}^{n}|Y_{1}^{n})-n\delta(\epsilon)\\
 & \overset{(d)}{\geq}\sum_{i=1}^{n}I(\psi_{1i};Y_{2i}|Y_{1i})-n\delta(\epsilon),\label{eq:H(J)2}
\end{align}
\end{subequations}where ($\mathit{a}$) follows because $h_{1}$
is a function of $J$ and $Y_{1}^{n}$; ($\mathit{b}$) follows by
the chain rule of mutual information and since mutual information
is non-negative; ($\mathit{c}$) follows by (\ref{eq:Fano}) and since
entropy is non-negative; and ($d$) follows by the chain rule for
entropy, since $Y_{2}^{n}$ and $Y_{1}^{n}$ are i.i.d., and due to
the fact conditioning decreases entropy. From (\ref{eq:H(J)}), (\ref{eq:H(J)1}),
and (\ref{eq:H(J)2}), we then have\begin{subequations}
\begin{align}
nR & \geq\sum_{i=1}^{n}I(\psi_{1i}\psi_{2i};X_{i}|Y_{2i})+I(\psi_{1i};Y_{2i}|Y_{1i})-n\delta(\epsilon)\label{ref-1}\\
 & \overset{(a)}{=}\sum_{i=1}^{n}I(X_{i};\psi_{1i}|Y_{1i})+I(X_{i};\psi_{2i}|Y_{2i}\psi_{1i})-n\delta(\epsilon),
\end{align}
\end{subequations}where ($\mathit{a}$) follows because of the Markov
chain relationship $(\psi_{1i},\psi_{2i})-X_{i}-Y_{2i}-Y_{1i}$, for
$i=1,...,n$. By defining $\hat{X}_{ji}=\psi_{ji}$ with $j=1,2$
and $i=1,...,n$, the proof is concluded as in {\cite{Steinberg}}.


\section*{Appendix B: Proof of Proposition \ref{prop:HB_Gauss}}

As explained in the text, we only need to focus on the case where
$D_{2}\leq D_{1}\leq\sigma_{x}^{2}$. As per the discussion in Appendix
A, we can assume, without loss of generality, that the Markov chain
relationship $X-Y_{2}-Y_{1}$ holds, so that \begin{subequations}\label{eqn:side_Gauss-1}
\begin{eqnarray}
Y_{2} & = & X+Z_{2}\\
\textrm{and }Y_{1} & = & Y_{2}+\tilde{Z}_{1},
\end{eqnarray}
\end{subequations}where $\tilde{Z}_{1}\sim\mathcal{N}(0,N_{1})$
is independent of $(X,Z_{2})$.

We first prove a converse. Calculating the rate-distortion function
in (\ref{eq:RHB1}) requires minimization over the pmf $p(\hat{x}_{1},\hat{x}_{2}|x)$
under the constraint (\ref{eq:const}). A minimizing $p(\hat{x}_{1},\hat{x}_{2}|x)$
exists by the Weierstrass theorem due to the continuity of the mutual
information and the compactness of the set of pmfs defined by the
constraint (\ref{eq:const})\cite{Bertsekas}. Fixing one such optimizing
$p(\hat{x}_{1},\hat{x}_{2}|x)$, the rate-distortion function (\ref{eq:RHB1})
can be written as
\begin{eqnarray}
R_{HB}^{CR}(D_{1},D_{2})= & I(X;\hat{X}_{2}|Y_{2})+I(\hat{X}_{1};Y_{2}|Y_{1}).\label{eq:1}
\end{eqnarray}
The first term in (\ref{eq:1}), i.e., $I(X;\hat{X}_{2}|Y_{2}),$
can be easily bounded using the approach in \cite[p. 5007]{Steinberg}.
Specifically, we have
\begin{eqnarray}
I(X;\hat{X}_{2}|Y_{2}) & = & h(X|Y_{2})-h(X|\hat{X}_{2}Y_{2})\nonumber \\
 & = & h(X|X+Z_{2})-h(X-\hat{X}_{2}|\hat{X}_{2},\hat{X}_{2}+(X-\hat{X}_{2})+Z_{2})\nonumber \\
 & = & h(X|X+Z_{2})-h(X-\hat{X}_{2}|\hat{X}_{2},(X-\hat{X}_{2})+Z_{2})\nonumber \\
 & \overset{(a)}{\geq} & h(X|X+Z_{2})-h(X-\hat{X}_{2}|(X-\hat{X}_{2})+Z_{2})\nonumber \\
 & \overset{(b)}{\geq} & \frac{1}{2}\log_{2}\left(2\pi e\frac{\sigma_{x}^{2}}{1+\frac{\sigma_{x}^{2}}{N_{2}}}\right)-\frac{1}{2}\log_{2}\left(2\pi e\frac{D_{2}}{1+\frac{D_{2}}{N_{2}}}\right)\nonumber \\
 & = & \frac{1}{2}\log_{2}\left(\frac{\sigma_{x}^{2}}{\sigma_{x}^{2}+N_{2}}\cdot\frac{D_{2}+N_{2}}{D_{2}}\right),\label{eq:Stein}
\end{eqnarray}
where ($a$) follows because conditioning decreases entropy; and ($b$)
follows from the maximum conditional entropy lemma \cite[p. 21]{Elgammal},
which implies that $h(E|E+Z_{2})\leq\frac{1}{2}\log_{2}(2\pi e\sigma_{E|E+Z_{2}}^{2})$
with $E=X-\hat{X}_{2}$. In fact, we have that $\sigma_{E|E+Z_{2}}^{2}\leq\frac{D_{2}}{1+\frac{D_{2}}{N_{2}}}$,
since the conditional variance $\sigma_{E|E+Z_{2}}^{2}$ is upper
bounded by the linear minimum mean square error of the estimate of
$E$ given $E+Z_{2}$. This mean square error is given by $\frac{D_{2}}{1+\frac{D_{2}}{N_{2}}}$,
since we have E$[E^{2}]\leq D_{2}$ and since $Z_{2}$ is independent
of $E$ due to the factorization (\ref{eq:joint}) and to the independence
of $X$ and $Z_{2}$. For the second term in (\ref{eq:1}), we instead
have the following:
\begin{eqnarray}
I(\hat{X}_{1};Y_{2}|Y_{1}) & = & h(Y_{2}|Y_{1})-h(Y_{2}|Y_{1},\hat{X}_{1})\notag\\
 & = & \frac{1}{2}\log_{2}\left(2\pi e\frac{N_{1}(N_{2}+\sigma_{x}^{2})}{N_{1}+N_{2}+\sigma_{x}^{2}}\right)-h(Y_{2}|Y_{1},\hat{X}_{1}).\label{2}
\end{eqnarray}
Moreover, we can evaluate
\begin{eqnarray}
h(Y_{2}|Y_{1},\hat{X}_{1}) & = & h(Y_{2},Y_{1}|\hat{X}_{1})-h(Y_{1}|\hat{X}_{1})\notag\\
 & = & h(Y_{2}|\hat{X}_{1})+h(Y_{1}|Y_{2},\hat{X}_{1})-h(Y_{1}|\hat{X}_{1})\notag\\
 & = & h(Y_{2}|\hat{X}_{1})-h(Y_{2}+\tilde{Z}_{1}|\hat{X}_{1})+h(Y_{2}+\tilde{Z}_{1}|Y_{2},\hat{X}_{1})\notag\\
 & \overset{(a)}{=} & h(Y_{2}|\hat{X}_{1})-h(Y_{2}+\tilde{Z}_{1}|\hat{X}_{1})+\frac{1}{2}\log_{2}(2\pi eN_{1}),\label{3}
\end{eqnarray}
where (\textit{a}) follows because $\tilde{Z}_{1}$ is independent
of $Y_{2}$ and of $\hat{X}_{1},$ due to the factorization (\ref{eq:joint})
and due to the independence of $\tilde{Z}_{1}$ and $X$. Next, we
obtain a lower bound on the term $h(Y_{2}+\tilde{Z}_{1}|\hat{X}_{1})$
in (\ref{3}) as a function of $h(Y_{2}|\hat{X}_{1})$ by using the
entropy power inequality (EPI) \cite[p. 22]{Elgammal}. Specifically,
by using the conditional version of EPI \cite[p. 22]{Elgammal}, we
have
\begin{eqnarray}
2^{2h(Y_{2}+\tilde{Z}_{1}|\hat{X}_{1})} & \geq & 2^{2h(Y_{2}|\hat{X}_{1})}+2^{2h(\tilde{Z}_{1}|\hat{X}_{1})}\notag\\
 & \overset{(a)}{=} & 2^{2h(Y_{2}|\hat{X}_{1})}+2^{2h(\tilde{Z}_{1})\notag}\\
 & = & 2^{2h(Y_{2}|\hat{X}_{1})}+2\pi eN_{1},\label{4}
\end{eqnarray}
where (\textit{a}) follows because $\tilde{Z}_{1}$ is independent
of $\hat{X}_{1}$ as explained above. The first two terms in (\ref{3})
can thus be bounded as
\begin{eqnarray}
h(Y_{2}|\hat{X}_{1})-h(Y_{2}+\tilde{Z}_{1}|\hat{X}_{1}) & \leq & h(Y_{2}|\hat{X}_{1})-\frac{1}{2}\log(2^{2h(Y_{2}|\hat{X}_{1})}+2\pi eN_{1})\notag\\
 & = & \frac{1}{2}\log_{2}\left(\frac{2^{2h(Y_{2}|\hat{X}_{1})}}{2^{2h(Y_{2}|\hat{X}_{1})}+2\pi eN_{1}}\right)\notag\\
 & \overset{(a)}{\leq} & \log_{2}\left(\frac{2\pi e(D_{1}+N_{2})}{2\pi e(D_{1}+N_{2})+2\pi eN_{1}}\right),\label{5}
\end{eqnarray}
where (a) follows because $\log_{2}\left(\frac{2^{2h(Y_{2}|\hat{X}_{1})}}{2^{2h(Y_{2}|\hat{X}_{1})}+2\pi eN_{1}}\right)$
is an increasing function of $h(Y_{2}|\hat{X}_{1})$ and $h(Y_{2}|\hat{X}_{1})\leq\frac{1}{2}\log_{2}(2\pi e(D_{1}+N_{2})),$
as can be proved by using the same approach used for the bounds (a)
and (b) in (\ref{eq:Stein})$.$ By substituting (\ref{5}) into (\ref{3}),
and using the result in (\ref{2}), we obtain
\begin{eqnarray}
I(\hat{X}_{1};Y_{2}|Y_{1}) & \geq & \frac{1}{2}\log_{2}\left(2\pi e\frac{N_{1}(N_{2}+\sigma_{x}^{2})}{N_{1}+N_{2}+\sigma_{x}^{2}}\right)-\frac{1}{2}\log_{2}\left(2\pi e\frac{N_{1}(D_{1}+N_{2})}{D_{1}+N_{2}+N_{1}}\right)\notag\\
 & = & \frac{1}{2}\log_{2}\left(\frac{(N_{2}+\sigma_{x}^{2})(D_{1}+N_{2}+N_{1})}{(N_{1}+N_{2}+\sigma_{x}^{2})(D_{1}+N_{2})}\right).\label{6}
\end{eqnarray}
 Finally, by substituting (\ref{eq:Stein}) and (\ref{6}) into (\ref{eq:1}),
we obtain the lower bound
\begin{eqnarray}
R_{HB}^{CR}(D_{1},D_{2}) & \geq & \frac{1}{2}\log_{2}\left(\frac{\sigma_{x}^{2}}{\sigma_{x}^{2}+N_{2}}\cdot\frac{D_{2}+N_{2}}{D_{2}}\right)+\frac{1}{2}\log_{2}\left(\frac{(N_{2}+\sigma_{x}^{2})(D_{1}+N_{2}+N_{1})}{(N_{1}+N_{2}+\sigma_{x}^{2})(D_{1}+N_{2})}\right)\notag\\
 & = & \frac{1}{2}\log_{2}\left(\frac{\sigma_{x}^{2}}{(\sigma_{x}^{2}+N_{1}+N_{2})}\cdot\frac{(D_{1}+N_{1}+N_{2})(D_{2}+N_{2})}{(D_{1}+N_{2})D_{2}}\right).\label{eq:converse}
\end{eqnarray}

For achievability, we calculate (\ref{eq:RHB1}) with $X=\hat{X}_{2}+Q_{2}$
and $\hat{X}_{2}=\hat{X}_{1}+Q_{1},$ where $Q_{1}\sim\mathcal{N}(0,D_{1}-D_{2})$
and $Q_{2}\sim\mathcal{N}(0,D_{2})$ are independent of each other
and of $(\hat{X}_{1},$$\tilde{Z}_{1},Z_{2})$. This leads to the
upper bound
\begin{eqnarray}
R_{HB}^{CR}(D_{1},D_{2}) & \leq & I(X;\hat{X}_{1}\hat{X}_{2}|Y_{2})+I(\hat{X}_{1};Y_{2}|Y_{1})\nonumber \\
 & = & I(X;\hat{X}_{2}|Y_{2})+I(\hat{X}_{1};Y_{2}|Y_{1})\nonumber \\
 & = & h(X|Y_{2})-h(X|Y_{2},\hat{X}_{2})+h(Y_{2}|Y_{1})-h(Y_{2}|Y_{1},\hat{X}_{1})\nonumber \\
 & = & h(X|X+Z_{2})-h(\hat{X}_{2}+Q_{2}|\hat{X}_{2}+Q_{2}+Z_{2},\hat{X}_{2})\nonumber \\
 &  & +h(X+Z_{2}|X+Z_{2}+\tilde{Z}_{1})-h(\hat{X}_{1}+Q_{1}+Q_{2}+Z_{2}|\hat{X}_{1}+Q_{1}+Q_{2}+Z_{2}+\tilde{Z}_{1},\hat{X}_{1})\nonumber \\
 & = & h(X|X+Z_{2})-h(Q_{2}|Q_{2}+Z_{2})+h(X+Z_{2}|X+Z_{2}+\tilde{Z}_{1})\nonumber \\
 &  & -h(Q_{1}+Q_{2}+Z_{2}|Q_{1}+Q_{2}+Z_{2}+\tilde{Z}_{1})\nonumber \\
 & \overset{(a)}{=} & \frac{1}{2}\log_{2}\left(2\pi e\frac{\sigma_{x}^{2}}{1+\frac{\sigma_{x}^{2}}{N_{2}}}\right)-\frac{1}{2}\log_{2}\left(2\pi e\frac{D_{2}}{1+\frac{D_{2}}{N_{2}}}\right)+\frac{1}{2}\log_{2}\left(2\pi e\frac{\sigma_{x}^{2}+N_{2}}{1+\frac{\sigma_{x}^{2}+N_{2}}{N_{1}}}\right)\nonumber \\
 &  & -\frac{1}{2}\log_{2}\left(2\pi e\frac{D_{1}+N_{2}}{1+\frac{D_{1}+N_{2}}{N_{1}}}\right)\nonumber \\
 & = & \frac{1}{2}\log_{2}\left(\frac{\sigma_{x}^{2}}{(\sigma_{x}^{2}+N_{1}+N_{2})}\cdot\frac{(D_{1}+N_{1}+N_{2})(D_{2}+N_{2})}{(D_{1}+N_{2})D_{2}}\right),\label{eq:achieve}
\end{eqnarray}
where ($a$) follows using $h(A|A+B)=\frac{1}{2}\log_{2}\left(2\pi e\frac{S_{A}}{1+\frac{S_{A}}{S_{B}}}\right),$
for $A$ and $B$ being independent Gaussian sources with $A\sim\mathcal{N}(0,S_{A})$
and $B\sim\mathcal{N}(0,S_{B})$. By comparing (\ref{eq:ext_converse})
with (\ref{eq:achieve}), we complete the proof.

\section*{Appendix C: Proof of (\ref{eq:RCR_Bin})}

Here, we prove that (\ref{eq:RCR}) equals (\ref{eq:RCR_Bin}) for
the given sources. For the converse, we have that 

\begin{eqnarray}
I(X;\hat{X}|Y) & = & H(X|Y)-H(X|\hat{X},Y)\nonumber \\
 & = & p-pH(X|\hat{X},Y\neq X)-(1-p_{2})H(X|\hat{X},Y=X)\nonumber \\
 & = & p-pH(X|\hat{X},Y\neq X)\nonumber \\
 & \overset{}{=} & p-pH(X|\hat{X})\nonumber \\
 & \overset{}{=} & p-pH(X\oplus\hat{X}|\hat{X})\nonumber \\
 & \overset{(a)}{\geq} & p-pH(X\oplus\hat{X})\nonumber \\
 & \geq & p-pH(D)\nonumber \\
 & = & p(1-H(D)),\label{eq:21}
\end{eqnarray}
where ($a$) follows because conditioning decreases entropy. Achievability
follows by calculating (\ref{eq:RCR}) with $X=\hat{X}\oplus Q$ where
$Q\sim\textrm{Ber}(D)$.

\section*{Appendix D: Proof of Proposition \ref{prop:HB_Bin}}

As explained in the text, we only need to focus on the case where
$D_{2}\leq D_{1}\leq1/2$. As for Appendix A and Appendix B, we can
assume, without loss of generality, that the joint pmf of $(x,y_{1},y_{2})$
factorizes as (\ref{eq:degraded_joint}) as shown Fig. \ref{fig:test}.
We first prove a converse. Similar to (\ref{eq:1}), we can write
the rate-distortion function (\ref{eq:RHB1}) as 
\begin{eqnarray}
R_{HB}^{CR}(D_{1},D_{2})= & I(X;\hat{X}_{2}|Y_{2})+I(\hat{X}_{1};Y_{2}|Y_{1}),\label{eq:10}
\end{eqnarray}
where the mutual information terms are calculated with a distribution
$p(\hat{x}_{1},\hat{x}_{2}|x)$ minimizing (\ref{eq:RHB1}) under
the constraint (\ref{eq:const}). The first term in (\ref{eq:10}),
i.e., $I(X;\hat{X}_{2}|Y_{2}),$ can be easily bounded by following
the same steps used in the derivation of (\ref{eq:21}), leading to
\begin{eqnarray}
I(X;\hat{X}_{2}|Y_{2}) & \geq & p_{2}(1-H(D_{2})).\label{eq:11}
\end{eqnarray}
For the second term in (\ref{eq:10}), we instead have the following:
\begin{eqnarray}
I(\hat{X}_{1};Y_{2}|Y_{1}) & = & H(Y_{2}|Y_{1})-H(Y_{2}|Y_{1},\hat{X}_{1})\notag\\
 & = & H(Y_{2}|Y_{1})-H(Y_{2},Y_{1}|\hat{X}_{1})+H(Y_{1}|\hat{X}_{1})\\
 & = & H(Y_{2}|Y_{1})-H(Y_{2}|\hat{X}_{1})-H(Y_{1}|\hat{X}_{1},Y_{2})+H(Y_{1}|\hat{X}_{1})\\
 & \overset{(a)}{=} & H(Y_{2}|Y_{1})-H(Y_{2}|\hat{X}_{1})-H(Y_{1}|Y_{2})+H(Y_{1}|\hat{X}_{1}),\label{eq:12}
\end{eqnarray}
where ($a$) follows because of the Markov chain condition $Y_{1}-Y_{2}-\hat{X}_{1}$.
The second term in the right-hand side of (\ref{eq:12}) can be evaluated
as 
\begin{eqnarray}
H(Y_{2}|\hat{X}_{1}) & = & H(Y_{2},X|\hat{X}_{1})-H(X|Y_{2},\hat{X}_{1})\nonumber \\
 & = & H(X|\hat{X}_{1})+H(Y_{2}|X,\hat{X}_{1})-H(X|Y_{2},\hat{X}_{1})\nonumber \\
 & = & H(X|\hat{X}_{1})+H(Y_{2}|X)-p_{2}H(X|Y_{2}\neq X,\hat{X}_{1})-(1-p_{2})H(X|Y_{2}=X,\hat{X}_{1})\nonumber \\
 & \overset{(a)}{=} & H(X|\hat{X}_{1})+H(p_{2})-p_{2}H(X|\hat{X}_{1})\nonumber \\
 & = & H(p_{2})+(1-p_{2})H(X|\hat{X}_{1})\label{eq:13}
\end{eqnarray}
where ($a$) follows because $H(Y_{2}|X)=H(p_{2})$. The fourth term
in the right-hand side of (\ref{eq:12}) can similarly be evaluated
as 
\begin{eqnarray}
H(Y_{1}|\hat{X}_{1}) & = & H(p_{1})+(1-p_{1})H(X|\hat{X}_{1}).\label{eq:14}
\end{eqnarray}
Substituting (\ref{eq:13}) and (\ref{eq:14}) in (\ref{eq:12}),
we obtain
\begin{eqnarray}
I(\hat{X}_{1};Y_{2}|Y_{1}) & = & H(p_{1})+(1-p_{1})H(X|\hat{X}_{1})-(H(p_{2})+(1-p_{2})H(X|\hat{X}_{1}))\nonumber \\
 &  & +H(Y_{2}|Y_{1})-H(Y_{1}|Y_{2})\nonumber \\
 & = & H(p_{1})-H(p_{2})-(p_{1}-p_{2})H(X|\hat{X}_{1})+H(Y_{2})-H(Y_{1})\nonumber \\
 & \overset{(a)}{\geq} & (p_{1}-p_{2})-(p_{1}-p_{2})H(D_{1})\label{eq:15}
\end{eqnarray}
where ($a$) follows since $H(Y_{2})=H(p_{2})+(1-p_{2})$ and $H(Y_{1})=H(p_{1})+(1-p_{1})$
and due to the inequality $H(X|\hat{X}_{1})\leq H(D_{1})$. Substituting
(\ref{eq:15}) and (\ref{eq:11}) into (\ref{eq:10}), we obtain
\begin{eqnarray}
R_{HB}^{CR}(D_{1},D_{2}) & \geq & p_{2}(1-H(D_{2}))+(p_{1}-p_{2})(1-H(D_{1}))\nonumber \\
 & = & p_{1}(1-H(D_{1}))+p_{2}(H(D_{1})-H(D_{2})).\label{eq:16}
\end{eqnarray}

For achievability, we calculate (\ref{eq:RHB1}) with $X=\hat{X}_{2}\oplus Q_{2}$
and $\hat{X}_{2}=\hat{X}_{1}\oplus Q_{1},$ where $Q_{1}\sim\textrm{Ber}(D_{1}*D_{2})$
and $Q_{2}\sim\textrm{Ber}(D_{2})$ are independent of each other
and of $(\hat{X}_{1},$$E_{1},E_{2})$ where $E_{j}=\boldsymbol{1}\{Y_{j}=\ce{e}\}$
for $j=1,2$. This leads to the upper bound
\begin{eqnarray}
R_{HB}^{CR}(D_{1},D_{2}) & \leq & I(X;\hat{X}_{1}\hat{X}_{2}|Y_{2})+I(\hat{X}_{1};Y_{2}|Y_{1})\nonumber \\
 & = & I(X;\hat{X}_{2}|Y_{2})+I(\hat{X}_{1};Y_{2}|Y_{1})\nonumber \\
 & = & H(X|Y_{2})-H(X|Y_{2},\hat{X}_{2})+H(Y_{2}|Y_{1})-H(Y_{2}|Y_{1},\hat{X}_{1})\nonumber \\
 & \overset{(a)}{=} & p_{2}-p_{2}H(X|\hat{X}_{2},Y_{2}\neq X)-(1-p_{2})H(X|\hat{X}_{2},Y_{2}=X)+p_{1}H\left(\frac{p_{2}}{p_{1}}\right)\nonumber \\
 &  & +\tilde{p}_{1}(1-p_{2})-p_{1}H(Y_{2}|\hat{X}_{1},Y_{1}=e)-(1-p_{1})H(Y_{2}|\hat{X}_{1},Y_{1}=X)\nonumber \\
 & \overset{(b)}{=} & p_{2}-p_{2}H(X|\hat{X}_{2})+p_{1}H\left(\frac{p_{2}}{p_{1}}\right)+\tilde{p}_{1}(1-p_{2})\nonumber \\
 &  & -p_{1}\left(H\left(\frac{p_{2}}{p_{1}}\right)+\frac{\tilde{p}_{1}(1-p_{2})}{p_{1}}H(X|\hat{X}_{1})\right)\nonumber \\
 & \overset{(c)}{=} & p_{2}-p_{2}H(\hat{X}_{2}\oplus Q_{2}|\hat{X}_{2})+\tilde{p}_{1}(1-p_{2})-\tilde{p}_{1}(1-p_{2})H(\hat{X}_{1}\oplus Q_{1}\oplus Q_{2}|\hat{X}_{1})\nonumber \\
 & \overset{(d)}{=} & p_{2}-p_{2}H(D_{2})+\tilde{p}_{1}(1-p_{2})-\tilde{p}_{1}(1-p_{2})H(D_{1})\nonumber \\
 & \overset{}{=} & p_{1}(1-H(D_{1}))+p_{2}(H(D_{1})-H(D_{2})),\label{eq:achieve-Bin}
\end{eqnarray}
where ($a$) follows because $H(Y_{2}|Y_{1})=p_{1}H\left(\frac{p_{2}}{p_{1}}\right)+\tilde{p}_{1}(1-p_{2})$;
($b$) follows because $H(Y_{2}|\hat{X}_{1},Y_{1}=X)=H(X|\hat{X}_{1},Y_{1}=X)=0$
and $H(Y_{2}|\hat{X}_{1},Y_{1}=e)=H\left(\frac{p_{2}}{p_{1}}\right)+\frac{\tilde{p}_{1}(1-p_{2})}{p_{1}}H(X|\hat{X}_{1})$;
($c$) follows by using the inverse test channels $X=\hat{X}_{2}\oplus Q_{2}$
and $\hat{X}_{2}=\hat{X}_{1}\oplus Q_{1}$; and ($d$) follows because
$Q_{2}\sim\textrm{Ber}(D_{2})$ and $Q_{1}\oplus Q_{2}\sim\textrm{Ber}(D_{1})$.
By comparing (\ref{eq:16}) with (\ref{eq:achieve-Bin}), we complete
the proof.

\section*{Appendix E: Proof of Proposition \ref{prop:cascade_Gauss}}

Here we provide the proof of Proposition \ref{prop:cascade_Gauss}.
To this end, we prove that for any pair $(D_{1},D_{2})$ there exists
a joint distribution $p(\hat{x}_{1},\hat{x}_{2}|x)$ such that (\ref{eq:const})
is satisfied and the conditions (\ref{eq:ach_R1}) and (\ref{eq:ach_R2})
coincide with (\ref{eq:converse_R1}) and (\ref{eq:converse_R2}),
respectively. This entails that the inner and outer bounds of Proposition
\ref{prop:outer} and Proposition \ref{prop:inner} coincide.

We distinguish the four region in the $(D_{1},D_{2})$ plane depicted
in Fig. \ref{fig:region}. If $D_{1}\geq\sigma_{x}^{2}\mbox{ and }D_{2}\geq\sigma_{x}^{2}$,
it is enough to set $\hat{X}_{1}=\hat{X}_{2}=0$ in (\ref{eqn: inner})
to prove. For $D_{1}\leq\sigma_{x}^{2}\mbox{ and }D_{2}\geq\min(D_{1},\sigma_{x}^{2})$,
we instead set $\hat{X}_{1}=\hat{X}_{2}$ and $X=\hat{X}_{1}+Q_{1}$
in (\ref{eqn: inner}), where $Q_{1}\sim\mathcal{N}(0,D_{1})$ is
independent of $\hat{X_{1}}$. Following the discussion in Sec. \ref{sub:Gaussian-HB},
it is easy to see that this choice is such that (\ref{eqn: inner})
coincides with (\ref{eqn: outer}). Next, in the sub-region where
$\mbox{\ensuremath{D_{1}}\ensuremath{\geq\sigma_{x}^{2}} }\mbox{and }D_{2}\leq\sigma_{x}^{2}$,
we select $\hat{X}_{1}=0$ and $X=\hat{X}_{2}+Q_{2}$ in (\ref{eqn: inner}),
where $Q_{2}\sim\mathcal{N}(0,D_{2})$ is independent of $\hat{X_{2}}$.
Finally, for the region in Fig. \ref{fig:region}, for which $D_{2}\leq D_{1}\leq\sigma_{x}^{2}$,
we choose $X=\hat{X}_{2}+Q_{2}$ and $\hat{X}_{2}=\hat{X}_{1}+Q_{1},$
where $Q_{1}\sim\mathcal{N}(0,D_{1}-D_{2})$ and $Q_{2}\sim\mathcal{N}(0,D_{2})$
are independent of each other and of $(\hat{X}_{1},E_{1},E_{2})$.
With this choice, following the derivations in Appendix B, we conclude
that condition (\ref{eq:ach_R1}) coincides with (\ref{eq:converse_R1}).
As for (\ref{eq:ach_R2}), we proceed as follows:
\begin{eqnarray}
I(X;\hat{X}_{1}|Y_{2})+I(X;\hat{X}_{2}|\hat{X}_{1}Y_{2}) & = & I(X;\hat{X}_{1}\hat{X}_{2}|Y_{2})\nonumber \\
 & = & h(X|Y_{2})-h(X|\hat{X}_{1},\hat{X}_{2},Y_{2})\nonumber \\
 & = & h(X|X+Z_{2})\nonumber \\
 &  & -h(\hat{X}_{1}+Q_{1}+Q_{2}|\hat{X}_{1},\hat{X}_{1}+Q_{1},\hat{X}_{1}+Q_{1}+Q_{2}+Z_{2})\nonumber \\
 & = & h(X|X+Z_{2})-h(Q_{1}+Q_{2}|Q_{1},Q_{1}+Q_{2}+Z_{2})\nonumber \\
 & = & h(X|X+Z_{2})-h(Q_{2}|Q_{2}+Z_{2})\nonumber \\
 & = & \frac{1}{2}\log_{2}\left(\frac{\sigma_{x}^{2}}{1+\frac{\sigma_{x}^{2}}{N_{2}}}\right)-\frac{1}{2}\log_{2}\left(\frac{D_{2}}{1+\frac{D_{2}}{N_{2}}}\right)\nonumber \\
 & = & \frac{1}{2}\log_{2}\left(\frac{\sigma_{x}^{2}}{\sigma_{x}^{2}+N_{2}}\frac{D_{2}+N_{2}}{D_{2}}\right)\nonumber \\
 & = & R_{\mathcal{G}}^{CR}(D_{2},N_{2}),
\end{eqnarray}
which concludes the proof.

\section*{Appendix F: Proof of Proposition \ref{prop:cascade_bin}}

Here we provide the proof of Proposition \ref{prop:cascade_bin}.
Following similar steps as in Appendix E, we prove that for any pair
$(D_{1},D_{2})$ there exists a joint distribution $p(\hat{x}_{1},\hat{x}_{2}|x)$
such that (\ref{eq:const}) is satisfied and the conditions (\ref{eq:ach_R1})
and (\ref{eq:ach_R2}) coincide with (\ref{eq:converse_R1}) and (\ref{eq:converse_R2}),
respectively. This entails that the inner and outer bounds of Proposition
\ref{prop:outer} and Proposition \ref{prop:inner} coincide.

We distinguish the four region in the $(D_{1},D_{2})$ plane depicted
in Fig. \ref{fig:region-bin}. If $D_{1}\geq1/2\mbox{ and }D_{2}\geq1/2$,
it is enough to set $\hat{X}_{1}=\hat{X}_{2}=0$ in (\ref{eqn: inner})
to prove the desired result. For $D_{1}\leq1/2\mbox{ and }D_{2}\geq\min(D_{1},1/2)$,
we instead set $\hat{X}_{1}=\hat{X}_{2}$ and $X=\hat{X}_{1}\oplus Q_{1}$
in (\ref{eqn: inner}), where $Q_{1}\sim\textrm{Ber}(D_{1})$ is independent
of $\hat{X}_{1}$. Following the discussion in Sec. \ref{sub:binary-HB},
it is easy to see that this choice is such that (\ref{eqn: inner})
coincides with (\ref{eqn: outer}). Next, in the sub-region where
$\mbox{\ensuremath{D_{1}\geq1/2}}\mbox{ and }D_{2}\leq1/2$, we select
$\hat{X}_{1}=0$ and $X=\hat{X}_{2}\oplus Q_{2}$ in (\ref{eqn: inner}),
where $Q_{2}\sim\textrm{Ber}(D_{2})$ is independent of $\hat{X_{2}}$.
Finally, for the region in Fig. \ref{fig:region-bin}, for which $D_{2}\leq D_{1}\leq1/2$,
we choose $X=\hat{X}_{2}\oplus Q_{2}$ and $\hat{X}_{2}=\hat{X}_{1}\oplus Q_{1},$
where $Q_{1}\sim\textrm{Ber}(D_{1}*D_{2})$ and $Q_{2}\sim\textrm{Ber}(D_{2})$
are independent of each other and of $(\hat{X}_{1},E_{1},E_{2})$.
With this choice, following the derivations in Appendix D, we conclude
that condition (\ref{eq:ach_R1}) coincides with (\ref{eq:converse_R1}).
As for (\ref{eq:ach_R2}), we proceed as follows:
\begin{eqnarray}
I(X;\hat{X}_{1}\hat{X}_{2}|Y_{2}) & = & H(X|Y_{2})-H(X|\hat{X}_{1},\hat{X}_{2},Y_{2})\nonumber \\
 & = & p_{2}-p_{2}H(X|\hat{X}_{1},\hat{X}_{2},Y_{2}\neq X)-(1-p_{2})H(X|\hat{X}_{1},\hat{X}_{2},Y_{2}=X)\nonumber \\
 & = & p_{2}-p_{2}H(X|\hat{X}_{1},\hat{X}_{2})\nonumber \\
 & \overset{(a)}{=} & p_{2}-p_{2}H(X|\hat{X}_{2})\nonumber \\
 & = & p_{2}-p_{2}H(\hat{X}_{2}\oplus Q_{2}|\hat{X}_{2})\nonumber \\
 & = & p_{2}-p_{2}H(D_{2})\nonumber \\
 & = & R_{B}^{CR}(D_{2},p_{2}),
\end{eqnarray}
where ($a$) follows by the Markov chain relationship $X-\hat{X}_{2}-\hat{X}_{1}$.
This completes the proof.

\section*{Appendix G: Proof of Proposition \ref{prop:HB_converse_ext}}

The proof of the achievability follows from standard arguments, similar
to \cite{HB}. For the converse, following the proof of \cite[Theorem 3]{HB}
we have that for any $(R,D_{e,1}+\epsilon,D_{e,2}+\epsilon,D_{1}+\epsilon,D_{2}+\epsilon)$
code, the following inequality holds:
\begin{align}
nR\geq\sum_{i=1}^{n}I(X_{i};U_{1i}|Y_{1i})+I(X_{i};U_{2i}|Y_{2i}),\label{eq:ext_converse}
\end{align}
with the definitions $U_{ji}\overset{\triangle}{=}(J,Y_{j}^{n\backslash i})$,
for $j=1,2$, with $Y_{j}^{n\backslash i}=[Y_{j1}^{i-1},Y_{j(i+1)}^{n}]$.
Note that with the given definition of $U_{ji}$ we have that the
$i$th element of the decoding functions (\ref{eq:dec1})-(\ref{eq:dec2})
can be written as $h_{ji}(J,Y_{j}^{n})=\hat{\textrm{x}}_{ji}(U_{ji},Y_{ji})$
for all $i=1,...,n$ and $j=1,2$. Now, defining $D_{e,ji}\overset{\triangle}{=}\textrm{E}[d_{e,j}(h_{ji}^{n}(M,Y_{j}^{n}),\psi_{ji}(X^{n})],$
we have the following chain of inequalities for the code at hand and
$j=1,2$: \begin{subequations}\label{eqn:expected}
\begin{eqnarray}
D_{e,ji} & = & \textrm{E\ensuremath{_{X^{n}Y_{j}^{n}}}}[d_{e,j}(h_{ji}^{n}(J,Y_{j}^{n}),\psi_{ji}(X^{n}))]\\
 & \overset{(a)}{=} & \textrm{E\ensuremath{_{X^{n}U_{ji}Y_{ji}}}}[d_{e,j}(\hat{\textrm{x}}_{ji}(U_{ji},Y_{ji}),\psi_{ji}(X^{n}))]\label{eq:}\\
 & \overset{}{=} & \textrm{\ensuremath{\textrm{E}_{X^{n}U_{ji}}\textrm{E}_{Y_{ji}}}}[d_{e,j}(\hat{\textrm{x}}_{ji}(U_{ji},Y_{ji}),\psi_{ji}(X_{i},X^{n\backslash i}))|X^{n}U_{ji}]\\
 & = & \sum_{x^{n}\in\mathcal{X}^{n},u_{ji}\in\mathcal{U}}^{n}p(x^{n},u_{ji})\textrm{E}_{Y_{ji}}\\
 &  & [d_{e,j}(\hat{\textrm{x}}_{ji}(U_{ji},Y_{ji}),\psi_{ji}(X_{i},X^{n\backslash i}))|X_{i}=x_{i},X^{n\backslash i}=x^{n\backslash i},U_{ji}=u_{ji}]\nonumber \\
 & \overset{(b)}{\geq} & \sum_{x^{n}\in\mathcal{X}^{n},u_{ji}\in\mathcal{U}}^{n}p(x^{n},u_{ji})\\
 &  & \textrm{\ensuremath{\textrm{E}_{Y_{ji}}}}[d_{e,j}(\hat{\textrm{x}}_{ji}(U_{ji},Y_{ji}),\psi_{ji}(X_{i},X^{n\backslash i}))|X_{i}=x_{i},X^{n\backslash i}=\textrm{x}{}^{*n\backslash i}(x_{i},u_{ji}),U_{ji}=u_{ji}]\nonumber \\
 & \overset{(c)}{=} & \sum_{x^{n}\in\mathcal{X}^{n},u_{ji}\in\mathcal{U}}^{n}p(x^{n},u_{ji})\textrm{\ensuremath{\textrm{E}_{Y_{ji}}}}[d_{e,j}(\hat{\textrm{x}}_{ji}(U_{ji},Y_{ji}),\hat{\textrm{x}}_{e,ji}(U_{ji},X_{i}))|X_{i}=x_{i},U_{ji}=u_{ji}]\\
 & \overset{}{=} & \textrm{E\ensuremath{_{X_{i}U_{ji}Y_{ji}}}}[d_{e,j}(\hat{\textrm{x}}_{ji}(U_{ji},Y_{ji}),\hat{\textrm{x}}_{e,ji}(U_{ji},X_{i}))],
\end{eqnarray}
\end{subequations}where ($a$) follows by using the definition of
random variables $U_{j}=(J,Y_{j}^{n\backslash i})$; ($b$) follows
by selecting $\textrm{x}{}^{*n\backslash i}(x_{i},u_{ji})$ as
\begin{align*}
\textrm{x}{}^{*n\backslash i}(x_{i},u_{ji}) & \in\textrm{\ensuremath{\textrm{argmin}_{x^{n\backslash i}\in\mathcal{X}^{n\backslash i}}}}\\
 & \textrm{\ensuremath{\textrm{E}_{Y_{ji}}}}[d_{e,j}(\hat{\textrm{x}}_{ji}(U_{ji},Y_{ji}),\psi_{ji}(X_{i},X_{i}^{n\backslash i}))|X_{i}=x_{i},X^{n\backslash i}=x^{n\backslash i},U_{ji}=u_{ji}];
\end{align*}
and ($c$) follows from the Markov chain relationship $Y_{ji}-(X_{i},U_{ji})-X^{n\backslash i}$
and from the definition $\hat{\textrm{x}}_{e,ji}(U_{ji},X_{i})=\psi_{ji}(X_{i},\textrm{x}{}^{*n\backslash i}(X_{i},U_{ji}))$.
Let $Q$ be a uniform random variable over the interval $[1,n]$ and
independent of the variables $(X^{n},Y_{1}^{n},Y_{2}^{n},U_{1}^{n},U_{2}^{n},\hat{X}_{1}^{n},\hat{X}_{2}^{n},\hat{X}_{e,1}^{n},\hat{X}_{e,2}^{n})$
and define the random variables $U_{j}\overset{\triangle}{=}(Q,U_{jQ}),$
$X\overset{\triangle}{=}X_{Q},$ $Y_{j}\overset{\triangle}{=}Y_{jQ},$
$\hat{X}_{j}\overset{\triangle}{=}\hat{X}_{jQ},$ and $\hat{X}_{e,j}\overset{\triangle}{=}\hat{X}_{e,jQ}$
for $j=1,2$. Moreover, note that $\hat{X}_{j}$ is a deterministic
function of $U_{ji}$ and $Y_{ji}$, and $\hat{X}_{e,j}$ is a deterministic
function of $U_{ji}$ and $X_{i}$ for $j=1,2$. The proof is completed
by using (\ref{eq:dist_const_ext}) and the fact that the term $I(X_{i};U_{1i}|Y_{1i})+I(X_{i};U_{2i}|Y_{2i})$
in (\ref{eq:ext_converse}) is convex with respect to the pmf $p(u_{1i},u_{2i}|x_{i})$,
using standard steps (see, e.g., \cite{Chia}).


\section*{Acknowledgment}

The work of O. Simeone is supported by the U.S. National Science Foundation
under grant CCF-0914899, and the work of H. V. Poor and R. Tandon
is supported in part by the U.S. Air Force Office of Scientific Research
under MURI Grant FA9550-09-1-0643 and in part by the U.S. National
Science Foundation under Grant CNS-09-05398.

\ifCLASSOPTIONcaptionsoff \newpage{}\fi

\end{document}